\documentclass[aps,print,superscriptaddress]{revtex4-1}

\usepackage{amsmath,amssymb,textcomp,calc,capt-of,ifthen}
\usepackage{times}
\usepackage{graphicx} 
\usepackage[english]{babel}

\begin{document}

%Title of paper
\title{Quantum erasure using entangled surface acoustic phonons}

\author{A. Bienfait}
\altaffiliation[Present address: ]{Universit\'e de Lyon, ENS de Lyon, Universit\'e Claude Bernard, CNRS, Laboratoire de Physique,
F-69342 Lyon, France}
\affiliation{Pritzker School of Molecular Engineering, University of Chicago, Chicago IL 60637, USA}
\author{Y. P. Zhong}
\affiliation{Pritzker School of Molecular Engineering, University of Chicago, Chicago IL 60637, USA}
\author{H.-S. Chang}
\affiliation{Pritzker School of Molecular Engineering, University of Chicago, Chicago IL 60637, USA}
\author{M.-H. Chou}
\affiliation{Pritzker School of Molecular Engineering, University of Chicago, Chicago IL 60637, USA}
\affiliation{Department of Physics, University of Chicago, Chicago IL 60637, USA}
\author{C. R. Conner}
\affiliation{Pritzker School of Molecular Engineering, University of Chicago, Chicago IL 60637, USA}
\author{\'E . Dumur}
\altaffiliation[Present address: ]{Universit\'e Grenoble Alpes, CEA, INAC-Pheliqs, 38000 Grenoble, France}
\affiliation{Pritzker School of Molecular Engineering, University of Chicago, Chicago IL 60637, USA}
\affiliation{Argonne National Laboratory, Argonne IL 60439, USA}
\author{J. Grebel}
\affiliation{Pritzker School of Molecular Engineering, University of Chicago, Chicago IL 60637, USA}
\author{G. A. Peairs}
\affiliation{Department of Physics, University of California, Santa Barbara CA 93106, USA}
\affiliation{Pritzker School of Molecular Engineering, University of Chicago, Chicago IL 60637, USA}
\author{R. G. Povey}
\affiliation{Pritzker School of Molecular Engineering, University of Chicago, Chicago IL 60637, USA}
\affiliation{Department of Physics, University of Chicago, Chicago IL 60637, USA}
\author{K. J. Satzinger}
\altaffiliation[Present address: ]{Google, Santa Barbara CA 93117, USA.}
\affiliation{Department of Physics, University of California, Santa Barbara CA 93106, USA}
\affiliation{Pritzker School of Molecular Engineering, University of Chicago, Chicago IL 60637, USA}
\author{A. N. Cleland}
\affiliation{Pritzker School of Molecular Engineering, University of Chicago, Chicago IL 60637, USA}
\affiliation{Argonne National Laboratory, Argonne IL 60439, USA}

\date{\today}

\begin{abstract}
Using the deterministic, on-demand generation of two entangled phonons, we demonstrate a quantum eraser protocol in a phononic interferometer where the which-path information can be heralded during the interference process. Omitting the heralding step yields a clear interference pattern in the interfering half-quanta pathways; including the heralding step suppresses this pattern. If we erase the heralded information \emph{after} the interference has been measured, the interference pattern is recovered, thereby implementing a delayed-choice quantum erasure. The test is implemented using a closed surface-acoustic-wave communication channel into which one superconducting qubit can emit itinerant phonons that the same or a second qubit can later re-capture. If the first qubit releases only half of a phonon, the system follows a superposition of paths during the phonon propagation: either an itinerant phonon is in the channel, or the first qubit remains in its excited state. These two paths are made to constructively or destructively interfere by changing the relative phase of the two intermediate states, resulting in a phase-dependent modulation of the first qubit's final state, following interaction with the half-phonon. A heralding mechanism is added to this construct, entangling a heralding phonon with the signalling phonon. The first qubit emits a phonon herald conditioned on the qubit being in its excited state, with no signaling phonon, and the second qubit catches this heralding phonon, storing which-path information which can either be read out, destroying the signaling phonon's self-interference, or erased.
\end{abstract}

%\keywords{}

\maketitle

\section{Introduction}

Quantum mechanics famously uses dual descriptions for quantum objects, representing these as waves or as particles depending on the situation. This is a manifestation of complementarity, and is central to understanding many interferometric experiments. The prototypical example is Young's two-slit experiment \cite{youngCourseLecturesNatural1807}: A wave description predicts an interference pattern, while a classical particle-based description, in which the path followed by the particle is known, shows no pattern.  For a quantum object passing through a two-path interferometer, an interference pattern is expected, but detecting which path the quantum follows changes this to a non-interfering particle-like description. Since the early days of quantum mechanics, many thought experiments (see e.g. \cite{bohrdiscussion,wheeler1978}) and their experimental realizations have tested the validity and domain of application of these orthogonal representations. These have led to the currently-accepted understanding that the wave or particle nature of a quantum remains undetermined until a measurement occurs.

Among these experiments, a quantum eraser scheme, as proposed by Scully and Dr\"{u}hl \cite{scullyQuantumEraserProposed1982}, investigates whether it is possible to undo the act of determining which path the quantum followed: is it possible to recover an interference pattern that was suppressed by acquisition of which-path information, by ``erasing'' that information?  This can be investigated using a three-step process: (1) observing an interference pattern in a two-path interferometer; (2) acquiring which-path information and observing the corresponding suppression of the interference; and (3) erasing the which-path information and recovering the interference pattern.  This test can further be combined with a version of Wheeler's delayed-choice test \cite{wheeler1978,wheeler1984}, where the act of recombining the paths of an interferometer occurs \emph{after} the quantum has entered the interferometer, thereby preventing the quantum from ``choosing'' a wave or particle nature before the superposition has been created. For a quantum eraser, in fact, the results should remain unchanged even if the acquisition and erasure of the which-path information occurs after the registration of the interferometric effect.

Realizations of quantum erasers have so far used photons, in both the optical and microwave bands. The first experimental realization used optical photons and marked the photon's propagation through a specific path by creating a path-specific polarization \cite{kwiatObservationQuantumEraser1992}. The first delayed-choice eraser test \cite{kimDelayedChoiceQuantum2000} triggered the emission of entangled photon pairs on each path of the interferometer, using one set of photons to complete the propagation through the interferometer, and the other set to mark and erase the which-path information. Further tests used setups where the marking of the which-path information and the interference detection took place at spatially distant locations, making the test robust to locality loopholes \cite{maQuantumErasureCausally2013}. More recently, a quantum eraser test using superconducting qubits and microwave photons was realized using a Ramsey interferometer, where the which-path information was acquired by coupling to an ancillary cavity \cite{liuTwofoldQuantumDelayedchoice2017}.

Here, we propose and implement a quantum eraser scheme using surface acoustic wave (SAW) phonons \cite{morgandavidSurfaceAcousticWave2007}. Building on a previously demonstrated interferometer \cite{bienfaitPhononmediatedQuantumState2019}, we implement the quantum erasure process by constructing a two-phonon entangled state, with the second phonon marking the which-path information. The slow propagation of this `herald' phonon is exploited to delay the which-path information detection \emph{after} detection of the result of the interferometric process, allowing for a delayed-choice quantum erasure.

\section{Proposal for quantum erasure via phonons}

Surface acoustic waves have now been proposed and used with a range of quantum systems \cite{delsing2019SurfaceAcoustic2019}, including the manipulation of electronic spins \cite{golterCouplingSurfaceAcoustic2016,whiteleySpinPhononInteractions2019}, microwave-to-optical photon transduction \cite{bochmann2013,vainsencherBidirectionalConversionMicrowave2016,shumeikoQuantumAcoustoopticTransducer2016}, and ferrying electrons between distant quantum dots \cite{mcneilOndemandSingleelectronTransfer2011,hermelinElectronsSurfingSound2011}. Superconducting qubits combined with standing-wave SAW devices \cite{gustafssonPropagatingPhononsCoupled2014,manentiCircuitQuantumAcoustodynamics2017,mooresCavityQuantumAcoustic2018,bolgarQuantumRegimeTwoDimensional2018,noguchiQubitAssistedTransductionDetection2017,satzingerQuantumControlSurface2018,slettenResolvingPhononFock2019,anderssonNonexponentialDecayGiant2019,ekstromPhononRoutingControlling2019} have allowed synthesis of arbitrary acoustic quantum states \cite{satzingerQuantumControlSurface2018} in the resonant coupling regime \cite{manentiCircuitQuantumAcoustodynamics2017,mooresCavityQuantumAcoustic2018,bolgarQuantumRegimeTwoDimensional2018,noguchiQubitAssistedTransductionDetection2017,satzingerQuantumControlSurface2018,slettenResolvingPhononFock2019,bienfaitPhononmediatedQuantumState2019,ekstromPhononRoutingControlling2019} as well as phonon-number resolved state detection in the dispersive regime \cite{slettenResolvingPhononFock2019}. Traveling-wave implementations have been used to emit and detect single-phonon SAWs \cite{gustafssonPropagatingPhononsCoupled2014}, route single phonons \cite{ekstromPhononRoutingControlling2019}, observe electromagnetically-induced transparency \cite{anderssonElectromagneticallyInducedTransparency2019}, as well as realize phonon-mediated quantum state transfer and remote entanglement \cite{bienfaitPhononmediatedQuantumState2019}.

\begin{figure}[htbp]
\includegraphics[width=8.6cm]{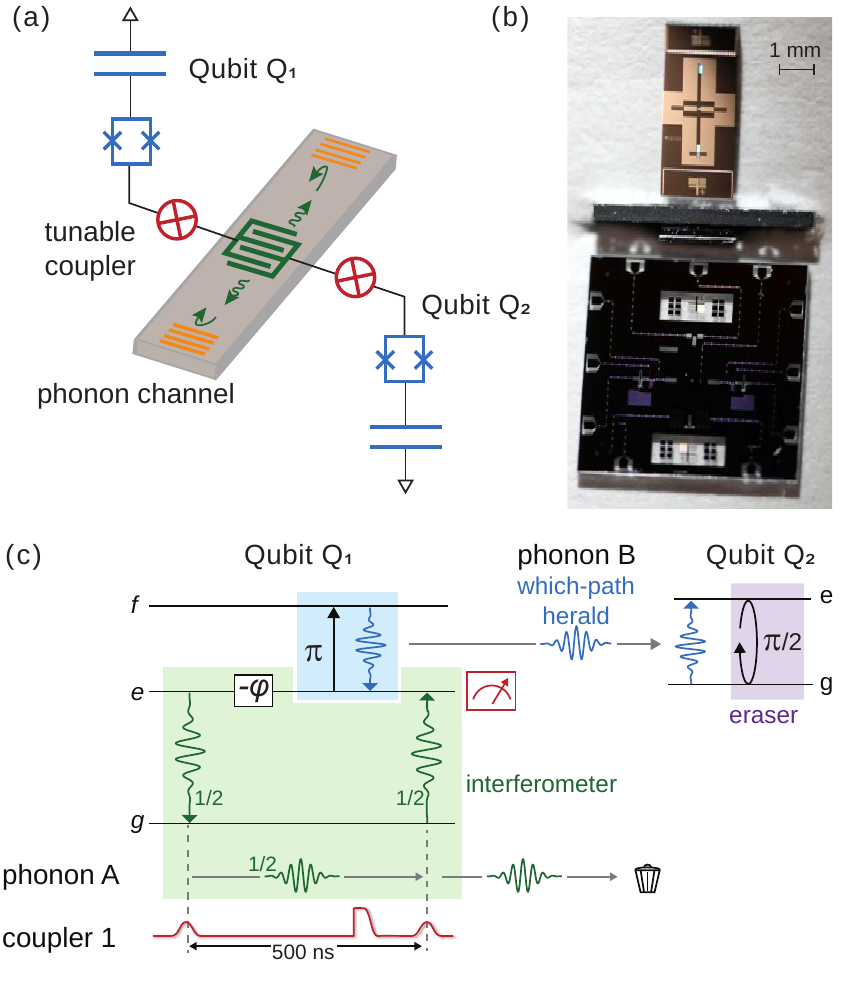}
\centering
\caption{\label{fig1}Experimental set-up and quantum eraser scheme.
(a) Two transmon superconducting qubits (blue) are coupled to a surface acoustic wave phononic channel (grey) via a central interdigitated transducer (IDT, green), using which both qubits can emit and capture itinerant phonons. The IDT is placed between two reflective mirror gratings (orange) that define a Fabry-P\'{e}rot cavity and reflect phonons within the mirrors' bandwidth back towards the IDT. Two tunable couplers (red) are used to dynamically control the coupling between the qubits and the IDT, allowing shaping the wavepackets of emitted phonons, and ensuring their efficient re-absorption after completing the 500~ns acoustic round-trip. The couplers also enable the controlled partial release of phonons.
(b) Optical micrograph of the device, showing (top) the acoustic Fabry-P\'{e}rot structure on a lithium niobate chip, (bottom) the two superconducting qubits and associated superconducting wiring on a separate sapphire chip, and (middle) a side-view of the flip-chip assembled device.
(c) When one of the qubits ($Q_1$) swaps a half-phonon ($A$) into the acoustic channel, an interferometer can be implemented (green box): once $A$ completes a round-trip within the acoustic cavity, its re-absorption probability by $Q_1$ depends on the relative phase accumulated by $Q_1$ and $A$, and leads to interference in $Q_1$'s final excitation probability. To implement a quantum eraser, we generate an entangled phonon herald marking the which-path information by generating a second, entangled phonon ($B$) conditionally on $Q_1$ being in $|e\rangle$ (blue box): this suppresses the interference. Capture and detection of the entangled herald $B$ by $Q_2$ acquires the which-path information after the interference of $Q_1$ and $A$ is complete, making this a time-delayed herald. Subsequently applying a $\pi/2$ pulse to $Q_2$ equalizes its $|g\rangle$ and $|e\rangle$ populations, erasing the which-path information and restoring the interference, thereby completing a delayed quantum eraser measurement.}
\end{figure}

The interferometry scheme we use for the quantum eraser protocol is described in Ref.~\cite{bienfaitPhononmediatedQuantumState2019}. The experimental layout of the device is shown in Fig.~\ref{fig1}a. Two nominally identical superconducting qubits \cite{kochChargeinsensitiveQubitDesign2007,barendsCoherentJosephsonQubit2013}, $Q_1$ and $Q_2$, are coupled via two tunable inductive couplers \cite{chenQubitArchitectureHigh2014} to a phonon channel comprising a central interdigitated transducer (IDT) located between two reflective mirror gratings. Each qubit can relax into this channel at a rate $\kappa(t)$, controlled by its tunable coupler,  emitting counter-propagating itinerant surface acoustic wave phonons via the IDT when the qubit is tuned near the IDT operating frequency of $\sim 4$~GHz. The two SAW mirrors, made of thin metallic gratings on either side of the IDT, ensure reflection of the phonons back towards the IDT when the phonons are in the mirrors' 125~MHz-wide operating bandwidth. Either qubit can efficiently re-absorb the itinerant phonons after the phonons complete a $\sim500$~ns-long round-trip: The tunable couplers' dynamic tuning is used to shape each emitted phonon wavepacket as well as to control their absorption \cite{zhongViolatingBellInequality2018}, enabling in theory their complete re-capture by either of the qubits \cite{korotkovFlyingMicrowaveQubits2011}. Experimentally, the qubit-to-qubit transfer efficiency is measured to be $\eta \sim 65$\%, limited by acoustic losses in the SAW device \cite{bienfaitPhononmediatedQuantumState2019}.

Here, we make use of the three lowest-energy qubit states, $|g\rangle$, $|e\rangle$ and $|f\rangle$. The qubits' anharmonicities $\chi/2\pi = (\omega_{ef}-\omega_{ge})/2\pi$ are respectively $-179$ and $-188$ MHz. The qubit intrinsic lifetimes are $T_1 = 18~\mu$s for both qubits, while the $g \textrm{-} e$ transition has a Ramsey $T_{2,ge,R}=1.2~\mu$s ($0.8~\mu$s) for $Q_1$ ($Q_2$), and  $T_{2,ef,R}=0.4~\mu$s for both qubits' $e \textrm{-} f$ transition. More details on the device and the phonon emission-capture protocol are available in \cite{bienfaitPhononmediatedQuantumState2019}.

A two-path interferometer can be realized in this device, shown in Fig.~\ref{fig1}b, by initializing one of the qubits (here $Q_1$) in its excited state and using its coupler to emit a half-phonon ($A$) with a symmetric wavepacket into the SAW channel. This results in the superposition state
\begin{equation}
    |\psi_1\rangle = (|e 0\rangle + |g 1\rangle)/\sqrt{2},
\end{equation}
writing $Q_1$'s state first and the phonon state second. Applying a detuning pulse on $Q_1$ of varying length introduces a relative phase $\varphi$ between the states $|e0\rangle$ and $|g1\rangle$ (defined here to be the phase accumulated by the phonon with respect to the qubit), yielding oscillations in the qubit occupancy after $Q_1$ re-captures the phonon\cite{bienfaitPhononmediatedQuantumState2019}. The origin of the interference can be understood by considering the outgoing acoustic field. This field has two contributions: the reflection of the incoming field combined with the field emitted by the qubit, whose population is also affected by the incoming field. External control of the qubit coupling rate $\kappa(t)$ ensures that the two contributions are equal in amplitude. The energy in the outgoing acoustic field thus only depends on the relative phase factor $e^{i \varphi}$. When $\varphi = 0$, absorption is the time-reversed emission process so that the qubit goes back to $|e\rangle$. The interference can be seen as destructive since the acoustic field reflected from the qubit acquires a $\pi$ phase shift and cancel out the acoustic field re-emitted by the qubit  and thus no phonon is re-emitted. When $\varphi = \pi$, the interference is constructive, and the qubit energy is transferred to the acoustic channel, leaving the qubit in $|g\rangle$; the re-emitted phonon eventually decays in the acoustic channel. The final state of the system can thus be written as a function of $\varphi$,
\begin{equation}
    |\psi_{f}\rangle = \frac{1+e^{i \varphi}}{2}|e 0\rangle +\frac{1-e^{i \varphi}}{2}  |g 1\rangle,
\end{equation}
resulting in the observation of an interference pattern in $Q_1$'s final excited state probability $P_e(t_{f})$ when sweeping the phase $\varphi$, with a period of $2 \pi$.

Two steps are required to realize a quantum eraser in this interferometer configuration. The first is to create which-path information, i.e. a herald indicating whether the qubit remained excited or instead phonon $A$ was emitted in the acoustic channel. Obtaining this information should result in the disappearance of the interference pattern, because this entangles the system under observation -- the qubit and traveling phonon $A$ -- with the measurement apparatus. The second step is to erase this knowledge, and look for a recovery of the interferometric pattern. Here, we use a protocol similar to that used in the original quantum eraser proposal \cite{scullyQuantumEraserProposed1982} as shown in Fig.~\ref{fig1}b. This protocol requires the on-demand generation of a second, entangled phonon to serve as a herald of the first, signalling phonon. Following the signaling half-phonon emission, we apply a transition-selective $\pi$ pulse on the $e \textrm{-} f$ transition of the qubit $Q_1$, then turn on the coupler, inducing $Q_1$ to emit a second phonon $B$ if initially in $|e\rangle$.  This phonon thereby heralds that the qubit is in its excited state (and that there is no $A$ phonon in the channel). Including the herald, the system state before re-absorption of phonon $A$ is then
\begin{equation}
    |\psi_2\rangle = \frac{1}{\sqrt{2}}|e 0\rangle |1\rangle_B + \frac{e^{i \varphi}}{\sqrt{2}} |g 1\rangle |0\rangle_B,
\end{equation}
displaying the entanglement of phonons $A$ and $B$. The entanglement of $Q_1$ with phonon $B$ makes the two states of the interferometer orthogonal, even after re-capture of phonon $A$, and prevents any interference. Phonon $B$ is then captured by qubit $Q_2$, putting $Q_2$ in $|e\rangle$ if $Q_1$ was in $|e\rangle$, transferring phonon $B$'s entanglement to $Q_2$ and thus placing the which-path information in $Q_2$ (this occurs after the interference has taken place, due to phonon $B$'s long (0.5~$\mu$s) transit time).

The which-path information can be erased by subsequently applying a $\pi/2$ pulse to $Q_2$, mapping $Q_2$'s state to a superposition of $|e\rangle$ and $|g\rangle$. For a particular phase choice for this $\pi/2$ erasure pulse, the final state of the system can be written as
\begin{equation}
    \begin{aligned}
        |\psi_f\rangle = &\frac{1+e^{i \varphi}}{\sqrt{8}} \left [ \lvert e0\rangle \lvert e\rangle - \lvert g1\rangle\lvert g\rangle\right]  \\
        + &\frac{1-e^{i \varphi}}{\sqrt{8}} \left[\lvert g1\rangle\lvert e\rangle - \lvert e0\rangle\lvert g\rangle \right ],
    \end{aligned}
    \label{eq.five}
\end{equation}
where qubit $Q_2$'s state is written last.

This expression shows that $Q_2$'s state remains entangled with the interferometer, but a measurement along its quantization axis no longer yields which-path information. The interference is therefore not directly recoverable by only measuring $Q_1$, but can be restored with a joint measurement of $Q_1$ and $Q_2$. This is similar to photon-based realizations of quantum eraser tests \cite{kwiatObservationQuantumEraser1992,kimDelayedChoiceQuantum2000,maQuantumErasureCausally2013,liuTwofoldQuantumDelayedchoice2017}, and the original quantum eraser proposal \cite{scullyQuantumEraserProposed1982}.

\section{Which-path herald}
Implementing the quantum eraser scheme hinges on our ability to emit a heralding phonon (phonon $B$) on $Q_1$'s $e\textrm{-}f$ transition, while preserving $Q_1$'s excited and ground-state populations. For a superconducting qubit coupled to a microwave environment, this can be achieved by either engineering the qubit's environment
\cite{gambettaSuperconductingQubitPurcell2011,jeffreyFastAccurateState2014,hoiProbingQuantumVacuum2015}, or manipulating the qubit's coupling to the environment \cite{pfaffControlledReleaseMultiphoton2017}. In our experiment, we make use of the former and harness the frequency-dependent response of the IDT \cite{morgandavidSurfaceAcousticWave2007,friskkockumDesigningFrequencydependentRelaxation2014,satzingerQuantumControlSurface2018,slettenResolvingPhononFock2019,anderssonElectromagneticallyInducedTransparency2019}. For a non-reflective uniform IDT of the type used here, the power conversion between microwave electrical and acoustic signals is proportional to the IDT conductance $G_{\textrm{a}}(\omega)$:
\begin{equation}
    G_{\textrm{a}}(\omega)/G_{\textrm{a}}(\omega_c) = [\sin(X)/X]^2,
\end{equation}
where $X = \pi N (\omega-\omega_c)/\omega_c$, $N=20$ is the number of  IDT finger pairs, $\omega_c = 2 \pi v/p$ the IDT central radial frequency, $p=0.985~\mu$m the IDT pitch and $v$ the SAW velocity within the IDT. The uniform profile of the IDT implies that $G_{\textrm{a}}=0$ for $X = \pm \pi$. At the corresponding frequencies $\omega_{\pm\pi}$, the qubit relaxation by phonon emission should be suppressed.

\begin{figure}[t]
\centering
\includegraphics[width=8.6cm]{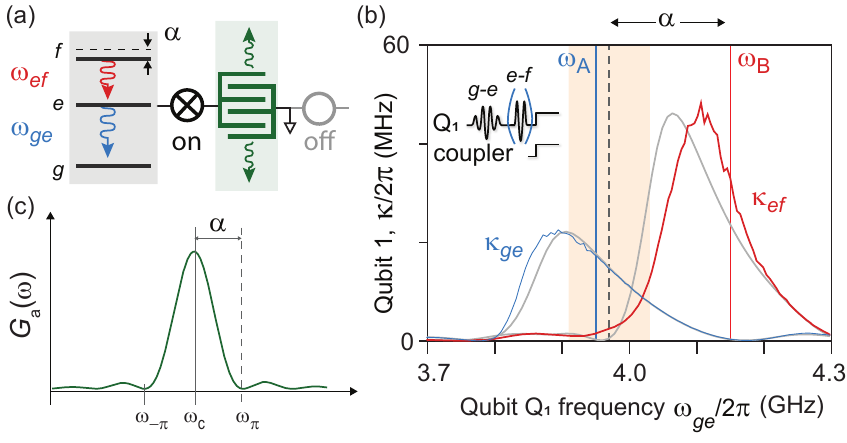}
\caption{\label{fig2} Single-qubit frequency- and state-dependent energy decay. (a) We monitor the decay of $Q_1$'s state after excitation respectively to $|e\rangle$ and $|f\rangle$ (pulse sequence is in inset of panel $b$), dominated by emission of phonons into the IDT. $Q_1$'s coupler is set to maximum coupling and $Q_2$'s coupler is turned off. (b) Fitting the population evolution (see \cite{SupplementaryMaterial}) enables us to extract the transition rate $\kappa_{ge}$ of transition $g \textrm{-} e$ (blue) and the transition rate $\kappa_{ef}$ of transition $e \textrm{-} f$ (red) as a function of $Q_1$ frequency. The frequency dependence of each transition rate is seen to follow the frequency-dependence of the IDT conductance (c). We identify two operating points $\omega_A$ and $\omega_B$. At frequency $\omega_{ge} = \omega_A$, phonon emission on the $g \textrm{-} e$ transition dominates, resulting in phonon emission at $\omega_{ge}/2\pi = 3.95$~GHz within the mirror bandwidth (orange), while decay on the $e\textrm{-}f$ transition is suppressed. Similarly, at frequency $\omega_{ge} = \omega_B = 2\pi\times 4.15$~GHz, phonon emission on the $e \textrm{-} f$ transition dominates, resulting in phonon emission at $\omega_{ef} = \omega_{ge}-\lvert\alpha\lvert = 2\pi \times 3.97$~GHz (grey dashed line), also within the mirror bandwidth, while decay on the $g\textrm{-}e$ transition is suppressed.}
\centering
\end{figure}

For this device, the qubit anharmonicity $\alpha$ is quite close to the difference between the IDT conductance minima at $\omega_{\pm\pi}$ and the IDT central frequency $\omega_{c}$. By tuning the qubit's $g \textrm{-} e$ emission frequency to $\omega_{ge} \sim \omega_{\pi}$, the $e \textrm{-} f$ transition is brought close to the IDT main emission peak, $\omega_{ef} \sim \omega_c$. Phonon emission on the $e \textrm{-} f$ transition is thus close to its maximum, while emission on the $g \textrm{-} e$ transition is heavily suppressed, making the proposed quantum eraser scheme possible. This is shown in Fig.~\ref{fig2}.

The $\kappa_{ge}(\omega_{ge})$ emission rate displays close to the expected behavior, as shown in Fig.~\ref{fig2}. Similarly, $\kappa_{ef}(\omega_{ge})$ also displays roughly the expected behavior: a shift in frequency by $\alpha$ compared to $\kappa_{ge}(\omega_{ge})$ and a factor of two increase in the rate ($\times 2.1 $ comparing the $\kappa_{ge}$ to the $\kappa_{ef}$ maxima), as expected for a weakly anharmonic qubit. The expected behavior, plotted as solid gray lines for both emissions in Fig.~\ref{fig2}b, is calculated from the qubit coupling to the IDT and the internal IDT frequency reflections, using an electrical model for the circuit and a coupling-of-modes model for the IDT \cite{morgandavidSurfaceAcousticWave2007}. These account for the nonlinearity of the qubit using ``black-box quantization'' \cite{niggBlackBoxSuperconductingCircuit2012, chenQubitArchitectureHigh2014}. The resulting modeled rates only account partially for the experimental results: while the agreement is satisfactory for the $e \textrm{-} g$ decay rates, we find a $50$~MHz misalignment in the modeled maximum of the $f \textrm{-} e$ decay compared to measurements. The modeling is explained in detail in \cite{SupplementaryMaterial}.

We extract two operating points, both within the IDT mirror bandwidth ($3.91$~GHz-$4.03$~GHz). At $\omega_{ge} = \omega_{A} = 2 \pi \times 3.95$~GHz, the $g \textrm{-} e$ emission time is $ 1/\kappa_{ge} = 9.3 \pm 0.1$~ns while the $e\textrm{-}f$ decay is suppressed by a relative factor $\kappa_{ge}/\kappa_{ef} = 5.9 \pm 0.1$, strongly favoring the emission of phonons on the $g \textrm{-} e$ transition.  When $\omega_{ge} = \omega_{B} = 2 \pi \times 4.15$~GHz, the emission time is $1/\kappa_{ef} = 4.8 \pm 0.1$~ns, with the phonon emitted at $\omega_{ef} = \omega_{B}-\lvert\alpha\lvert = 2 \pi \times 3.97$~GHz while the decay on $g \textrm{-} e$ is suppressed by a factor $\kappa_{ef}/\kappa_{ge} = 84 \pm 3$: this is the operating point for emitting the which-path herald.

\begin{figure}[t]
\centering
\includegraphics[width=8.6cm]{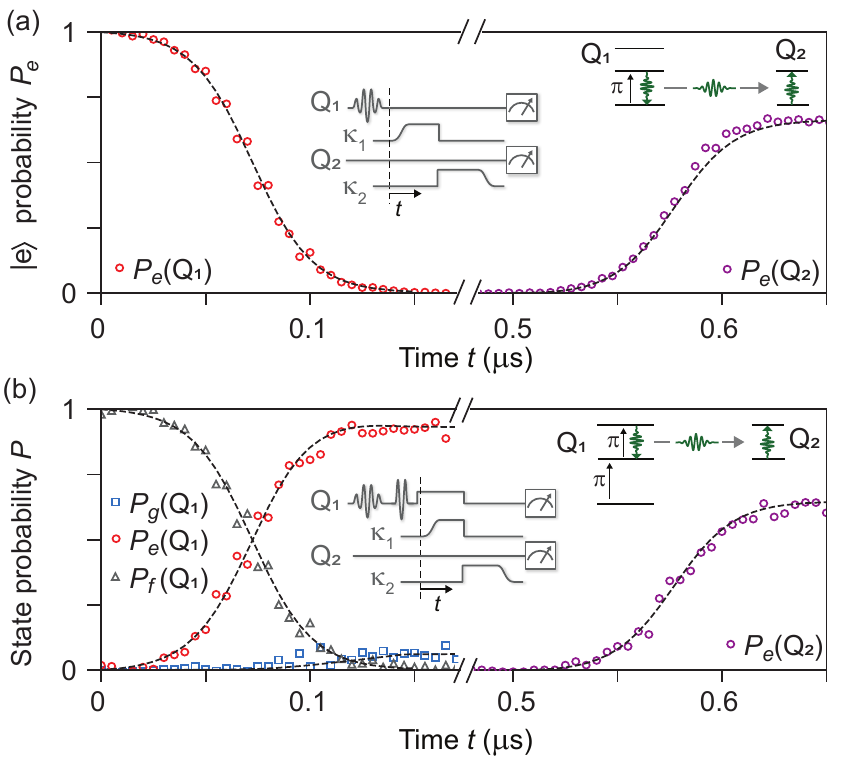}
\caption{\label{fig3}Which-path heralding. (a) After exciting $Q_1$ to $|e\rangle$ with $\omega_{ge} = 2 \pi \times 3.95$~GHz, $Q_1$'s tunable coupling $\kappa_1$ is modulated dynamically to release a symmetric phonon wavepacket with characteristic time $1/\kappa=15$~ns on $Q_1$'s $g \textrm{-} e$ transition. The emitted phonon is later captured by $Q_2$ on its $g \textrm{-} e$ transition.  (b) Inset pulse sequence: We initialize $Q_1$ to $|f\rangle$ using two sequential pulses at the $g \textrm{-} e$ and $e \textrm{-} f$ transition, with $\omega_{ef} = \omega_{ge}-\lvert\alpha\lvert = 2 \pi \times 3.97$~GHz. We then modulate $\kappa_1$ to emit a phonon on $Q_1$'s $e \textrm{-} f$ transition. The emitted phonon is later captured by $Q_2$ on its $g \textrm{-} e$ transition. During this process, $Q_1$'s $|g\rangle$ population increases from 0 to 0.06 due to spurious relaxation from $|e\rangle$ to $|g\rangle$. Insets show the pulse sequences and schematics of the expected transfers. Open symbols represent the qubits' populations measured at time $t$; dashed lines correspond to a numerical model taking into account the qubits' decoherence and phonon losses.}
\centering
\end{figure}

Operating at frequency $\omega_A$, we use the tunable couplers to efficiently shape the emitted and absorbed wave-packets, see \cite{korotkovFlyingMicrowaveQubits2011,campagne-ibarcqDeterministicRemoteEntanglement2018,kurpiersDeterministicQuantumState2018, axlineOndemandQuantumState2018, zhongViolatingBellInequality2018, bienfaitPhononmediatedQuantumState2019, SupplementaryMaterial}. The couplers are controlled so the emitted wavepackets have a cosecant shape with characteristic time $1/\kappa_c = 15$~ns \cite{SupplementaryMaterial}. In Fig.~\ref{fig3}a, we measure the transfer efficiency by emitting one phonon using $Q_1$'s $g \textrm{-} e$ transition and capturing it later using $Q_2$'s $g \textrm{-} e$ transition, with an efficiency $\eta_A = P_{2e}(t_f)/P_{1e}(0)  = 0.66 \pm 0.01$, limited by acoustic losses \cite{bienfaitPhononmediatedQuantumState2019,SupplementaryMaterial}. The same operation realized using  $Q_1$'s $e \textrm{-} f$ transition (Fig.~\ref{fig3}b) while operating at frequency $\omega_B$ yields the same efficiency, $\eta_B = P_{2e}(t_f)/P_{1f}(0)  = 0.64 \pm 0.02$. Due to the imperfectly-suppressed $1/\kappa_{ge} = 0.4 ~\mu$s decay, a small population is transferred from $|e\rangle$ to $|g\rangle$ during this process, leading to $P_{1g}(t_f) = 0.06 \pm 0.02$. As a consequence, exciting and then emitting a phonon on the $e \textrm{-} f$ transition to herald the which-path information will have at most a $\eta_h = 94 \pm 2$\% efficiency due to this spurious decay. The probability of actually detecting this information is limited to $\eta_B$.

\begin{figure*}[t]
\centering
\includegraphics[width=17.6cm]{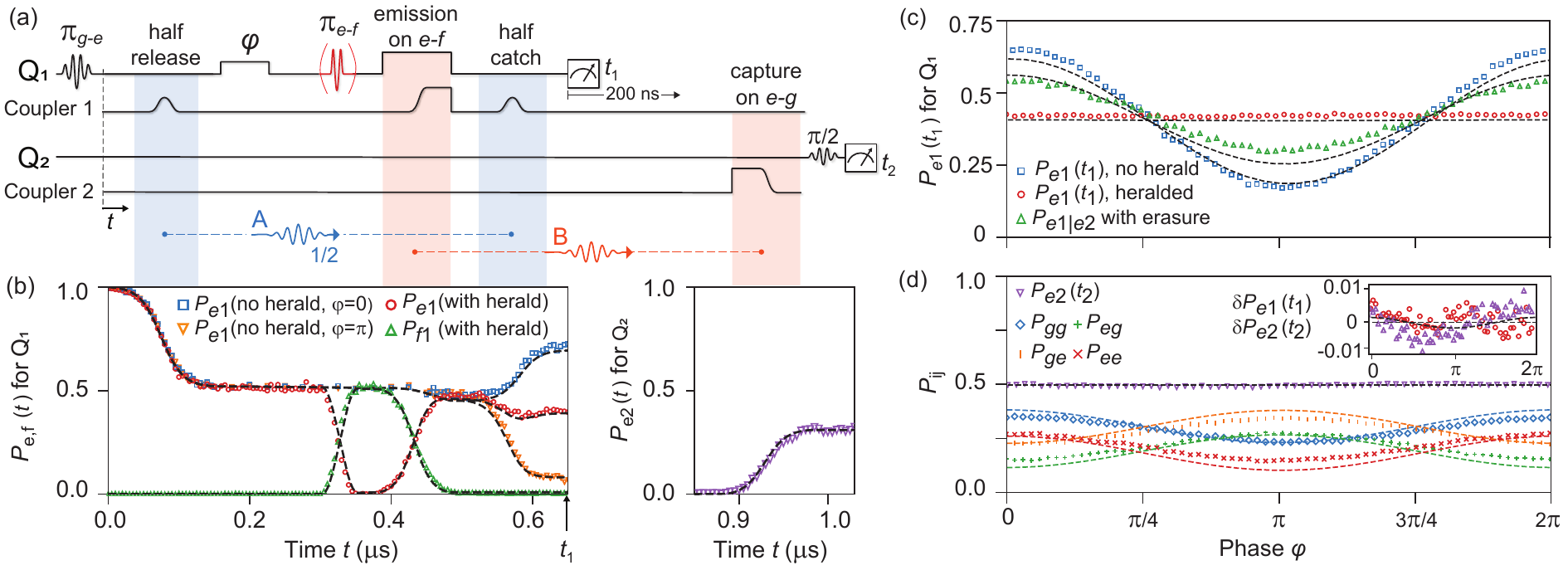}
\caption{\label{fig4} Quantum eraser. (a) Pulse sequence: With $Q_1$ in $|e\rangle$, its coupler is used to half-release a phonon at $\omega_A$ (blue). $Q_1$'s frequency is then detuned, accumulating a phase $\varphi$ between the half-phonon and $Q_1$'s $|e\rangle$ state. An optional $\pi$ pulse on $Q_1$'s $e \textrm{-} f$ transition (red) is followed by the coupler-controlled emission of a phonon at $\omega_B$ (orange), heralding that $Q_1$ is in $|e\rangle$ and returning $Q_1$ to $|e\rangle$. Following the optional heralding, $Q_1$ catches the half-phonon (blue) and is measured, completing the interferometry. Following $Q_1$'s measurement, $Q_2$ catches the optional heralding phonon (orange) and is measured at time  $t_{2} = 1.1~\mu\textrm{s}$. (b) Left: $Q_1$'s $|e\rangle$ and $|f\rangle$ state populations as a function of time $t$, showing the unheralded $P_e(t)$ for $\varphi=0$ and $\varphi=\pi$, which at time $t_{1}$ displays the interference maximum and minimum, and for the heralded $P_e$, which at time $t_{1}$ does not have a $\varphi$ dependence. Also shown is $Q_1$'s $|f\rangle$ state population when the herald is generated. Right: $Q_2$'s excited state $P_e(t)$ when the herald is generated, which ideally would reach the value $1/2$ but is limited by acoustic losses to $\eta_b \times 0.5 \approx 0.32$. (c) Interference fringes $P_e(\varphi)$ are visible when the herald is absent, but disappear when the herald reports which-path information ($Q_1$ in $|e\rangle$). If the herald is generated but the information in $Q_2$ is erased, by applying a $\pi/2$ pulse on $Q_2$'s $g \textrm{-} e$ transition, the fringes reappear when $Q_1$'s measurement is conditioned on measuring $Q_2$ in $|e\rangle$. This occurs even though $Q_1$'s measurement was already complete by the time the information in $Q_2$ is erased. (d) Probability of measuring $Q_2$ in $|e\rangle$, showing lack of dependence on $\varphi$. Also shown are variations in two-qubit probabilities $P_{gg}$, $P_{ge}$, $P_{eg}$ and $P_{ee}$. Inset shows the lack of variation of both qubits' $|e\rangle$-state probabilities $P_{e1}$ and $P_{e2}$ as a function of $\varphi$ when applying the $\pi_{ef}$ pulse. Dashed lines in all panels correspond to a numerical model taking into account the qubits' decoherence and phonon losses, see \cite{SupplementaryMaterial}.}
\centering
\end{figure*}

\section{Quantum erasure implementation}

We implement the full quantum eraser scheme as shown in Fig.~\ref{fig4}. First, we demonstrate single-phonon interferometry without heralding: Qubit $Q_1$, initialized in $|e\rangle$, emits, and later re-captures, a half-phonon on its $g \textrm{-} e$ transition at $\omega_A$. Following release, a detuning pulse applied to $Q_1$ accumulates a phase $\varphi$ between the traveling half-phonon and $Q_1$ (pulse sequence in panel a; intermediate measurements in panel b). This results in an interference pattern in the final excitation probability $P_{e1}(t_{1} = 650~\textrm{ns})$ of $Q_1$ as a function of $\varphi$ (panel c). The oscillations have an average occupation of $0.41$ with peak-to-peak amplitude $0.49$. These are reduced from the ideal values of $1/2$ and $1$ due to acoustic losses, $Q_1$ decoherence, and the finite readout visibility. Taking these effects into account, a numerical model (see \cite{SupplementaryMaterial}) provides similar results (panel c).

A which-path herald is generated by inserting an intermediate $\pi$ pulse on $Q_1$'s $e \textrm{-} f$ transition followed by emission of a phonon at $\omega_B$ on $Q_1$'s $e \textrm{-} f$ transition, returning $Q_1$ to $|e\rangle$ (see panels Fig.~\ref{fig4}a and b). Generating the herald destroys the interference pattern, as expected. The amplitude in the heralded $P_{e1}$ displays small fluctuations with amplitude $\sim 0.01$. This could be attributed to the imperfect information acquisition discussed in Fig.~\ref{fig3}, with our model shown by the dashed line, but is below the noise threshold. We note that even if the heralding phonon is not captured and detected via $Q_2$, the interference is not recovered, as $Q_1$'s state is now irremediably coupled to the herald and thus to the environment.

The final step of the quantum eraser test is to erase the heralded information, and thereby recover the interference pattern. As the heralding phonon marks whether $Q_1$ was in $|e\rangle$, its capture using $Q_2$ followed by a $\pi/2$ pulse on $Q_2$'s $g \textrm{-} e$ transition erases the information that could distinguish the two paths. This erasure can be performed in a time-delayed manner by capturing the herald and measuring $Q_2$ after the measurement of $Q_1$. We thus implement the measurement of $Q_1$ immediately following its interaction with the returning half-phonon, completing the interferometry, and before absorbing and detecting the herald using $Q_2$. This requirement limits $Q_1$'s readout time to 200~ns, decreasing its readout visibility from 96\% to 81\%.

As $Q_2$ is still entangled with the interferometer, simply tracing out $Q_2$'s state (equivalently, not measuring $Q_2$) will not recover the interference pattern; instead, we must condition the measurements of $Q_1$ on measurements of $Q_2$, even though measuring $Q_2$ does not yield any heralded information (see Eq.~(\ref{eq.five})). In Fig.~\ref{fig4}d, we plot all joint qubit probabilities as a function of $\varphi$: all have an oscillation pattern of amplitude $0.12$, while the excitation probabilities $P_{e1}$, $P_{e2}$ for each qubit evaluated separately only display very weak oscillations, below 1\%. To make a fair comparison with the original interference pattern, we next consider the conditional measurement $P_{e1|e2} = P_{ee}/(P_{ge}+P_{ee})$, the probability of measuring $Q_1$ in $|e\rangle$ conditioned on $Q_2$ being measured in $|e\rangle$. This probability has a mean identical to that measured without a herald, but the amplitude of the oscillations is reduced by 48\%, due to the inefficient capture of the second phonon and thus an incomplete erasure of information, as well as the additional decoherence in $Q_2$.

A model taking into account these losses and $Q_2$'s finite coherence time partially accounts for the amplitude reduction, as shown by the dashed line. We attribute the remaining discrepancy to decoherence occurring during the measurement, which we have not taken into account.

In conclusion, we have successfully completed a quantum eraser protocol, using an acoustic Fabry-P\'{e}rot interferometer. We realized three distinct steps in this process, first observing an interferogram; next, marking the which-path information which makes the interference fringes disappear, and third, erasing the which-path information which leads to the recovery of an interference signal. The erasure of the which-path information occurs after registering the result of the interference, making this a delayed-choice quantum eraser. The which-path detection was implemented by signaling using a heralding phonon.

This construct enabled us to demonstrate and exploit a two-phonon entanglement, opening the door to two-phonon interferometry, acoustic Bell tests \cite{fransonBellInequalityPosition1989} and phonon coherence length measurements \cite{rarityTwophotonInterferenceMachZehnder1990}. Phonon heralding as demonstrated here could also be used to mitigate propagation losses in future acoustic experiments and implement for example high-fidelity acoustic quantum state transfer  and remote entanglement, using schemes analogous to Refs. \cite{bernienHeraldedEntanglementSolidstate2013,kurpiersQuantumCommunicationTimebin2019}.

\begin{acknowledgments}
The authors thank  P. J. Duda and K. M\o lmer for helpful discussions and thank W.D. Oliver and G. Calusine at Lincoln Laboratories for the provision of a traveling-wave parametric amplifier (TWPA). Devices and experiments were supported by the Air Force Office of Scientific Research and the Army Research Laboratory. K.J.S. was supported by NSF GRFP (NSF DGE-1144085),
\'E.D. was supported by LDRD funds from Argonne National Laboratory, and A.N.C. was supported by the DOE, Office of Basic Energy Sciences. This work was partially supported by the UChicago MRSEC (NSF DMR-1420709) and made use of the Pritzker Nanofabrication Facility, which receives support from SHyNE, a node of the National Science Foundation's National Nanotechnology Coordinated Infrastructure (NSF NNCI-1542205). The authors declare no competing financial interests. The datasets supporting this work are available from the corresponding author on request. Correspondence and requests for materials should be addressed to A.~N.~Cleland (anc@uchicago.edu).
\end{acknowledgments}

\clearpage

\begin{center}
\textbf{\Large{Supplementary Materials for Quantum erasure using entangled surface acoustic phonons}}
\end{center}

\setcounter{figure}{0} 
\setcounter{section}{0}    
\renewcommand{\thefigure}{S\arabic{figure}}
\renewcommand{\thetable}{S\arabic{table}}
\renewcommand{\theequation}{S\arabic{equation}}

\section{Device, experimental setup and techniques}

\begin{figure}[b]
\centering
\includegraphics{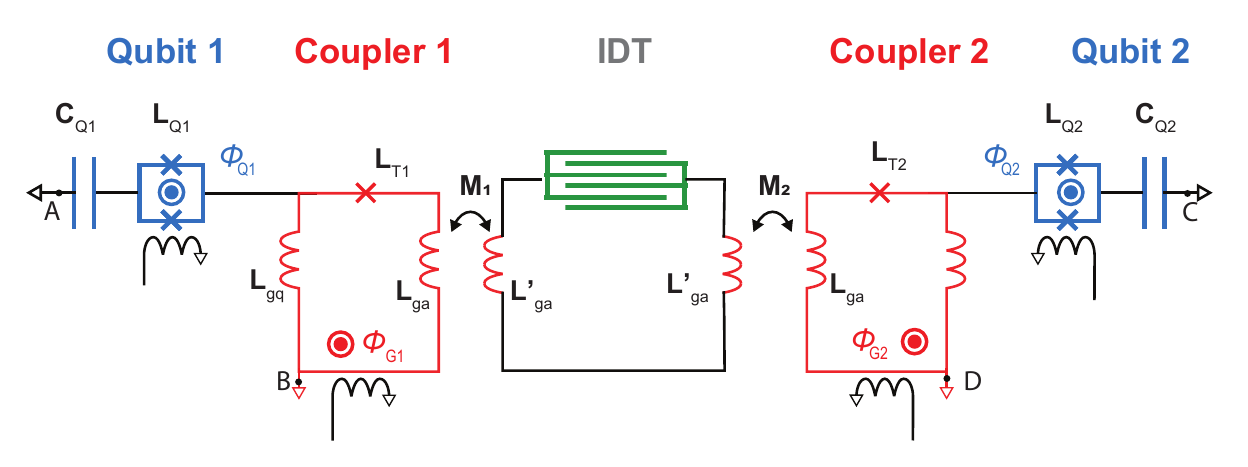}
\caption{\label{schema} Electrical circuit. Elements in blue are the qubit equivalent circuits, in red the variable couplers and the inductive couplers between the qubit sapphire chip and the acoustic lithium niobate chip, and green the interdigitated transducer (IDT) for phonon emission and capture.}
\centering
\end{figure}

The flip-chip device, setup and techniques used for this experiment are strictly identical to \cite{bienfaitPhononmediatedQuantumState2019},  except that the data shown in this paper were acquired in a separate cool-down of the cryostat used for the experiment (base temperature $<7$~mK). A full wiring diagram and a description of the room-temperature set-up may be found in Ref. \cite{zhongViolatingBellInequality2018}. The fabrication description is given in Ref. \cite{satzingerQuantumControlSurface2018}. The circuit is shown in Fig.~S1. Compared to Ref. \cite{bienfaitPhononmediatedQuantumState2019}, we note a 5\% shift in the nominal values of the Josephson junctions of the two tunable couplers, as well as an overall reduction of the coherence times of the qubits.

For this run, we implemented in addition a three-state dispersive readout. Each qubit readout resonator is a $\lambda/4$ resonator inductively coupled to a $\lambda/2$ Purcell filter. A 500-ns microwave tone is applied at resonance with each qubit readout resonator and the transmitted signal is successively amplified by a traveling-wave parametric amplifier \cite{macklinQuantumlimitedJosephsonTravelingwave2015}, a high-electron mobility transistor amplifier, and a room-temperature amplifier, before homodyne mixing and recording the integrated value of the quadrature amplitudes $I$ and $Q$. To estimate the fidelity of the preparation and readout of each state, we successively prepare each qubit in $|g\rangle$, $|e\rangle$ or $|f\rangle$ and repeat each measurement 4000 times. The state-dependent dispersive shift of the readout resonator allows us to attribute a sector of the $IQ$ plane to each state, enabling us to identify the qubit state from any single-shot readout based on its recorded $I$ and $Q$ values. These calibrations also determine the fidelity of each state readout, which are all above 90\%, see Table S1. Data shown in Fig.~3 and Fig.~4b in the main text are corrected for readout errors using this calibration.

To perform a delayed-choice quantum eraser test, we modified the $Q_1$ measurement procedure to fit within a phonon round-trip time, by shortening its readout pulse from 500 ns to 200 ns. When performing a two-state readout, this decreases the visibility of $Q_1$'s $|e\rangle$ and $|g\rangle$ states to 81\%.

\begin{figure}[tb]
\centering
\includegraphics[width=0.95\textwidth]{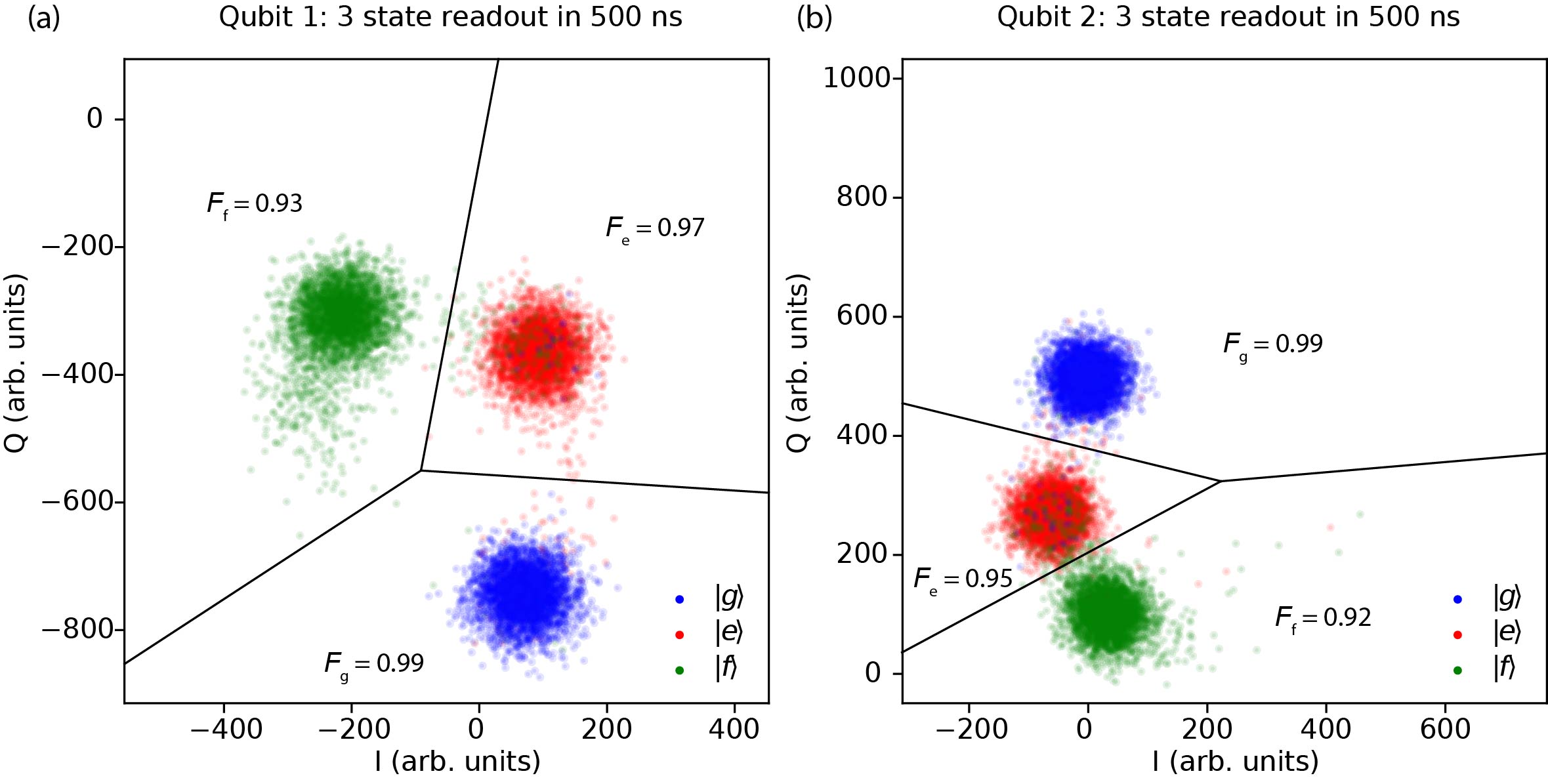}
\caption{\label{readout} Qubits $Q_1$ (a) and $Q_2$ (b) single-shot readout using a 500-ns readout pulse. Dots indicate the coordinates in the IQ-plane of each integrated  integrated readout pulses for the 4000 measurements realized after preparing each qubit either in $|g\rangle$ (blue), $|e\rangle$ (red) or $|f\rangle$ (green). This calibration allows us to assign any given measurement to the ground, excited or second excited state, as separated by the black lines in the IQ plane. Corresponding fidelities are given in inset.}
\centering
\end{figure}

\section{Relaxation rates and circuit modeling}

In this section, we describe the modeling and the measurements of the relaxation rates when one qubit ($Q_1$) is maximally coupled to the IDT, and the other qubit is disconnected (coupler 2 turned off), see Fig.~2 in the main text.  For a given qubit frequency, the operating points are determined by (1) maximizing the coupler-induced frequency shift on the qubit, (2) maximizing the other qubit relaxation time. We prepare the qubit in $|e\rangle$ or $|f\rangle$ by the successive application of resonant $\pi$ pulses and measure the qubit state populations after a varying amount of time $t$ during which coupler 1 is open. The measurements realized on $Q_1$ for the two operating points described in the main text are shown in Fig.~S3.

A weakly anharmonic transmon- or xmon-style qubit is expected to have decay rates very similar to a harmonic oscillator \cite{kochChargeinsensitiveQubitDesign2007}, with the population of the $|e\rangle$ and $|f\rangle$ excited states evolving as
\begin{eqnarray}
\dot{P}_f &=& -\kappa_{ef} P_f - \kappa_{gf} P_f\\
\dot{P}_e &=& -\kappa_{ge} P_e + \kappa_{ef} P_f, \label{T1Formula}
\end{eqnarray}
where $\kappa_{ef} = 2 \kappa_{ge}$ and $\kappa_{gf}=0$. Here, due to the IDT response, we measure a very different behavior.

We first make the assumption that $\kappa_{gf}=0$, as this two-phonon relaxation is expected to be exponentially suppressed for a transmon \cite{kochChargeinsensitiveQubitDesign2007}. To determine the rates $\kappa_{ge}$ and $\kappa_{ef}$, we start by fitting the decay from $|e\rangle$ after excitation to $|e\rangle$ with a single decaying exponential for all qubit frequencies. We only consider times past the transient on-set of the coupler ($t \geq 3$~ns) and prior to any re-excitation of the qubit by the phonons reflected off the mirrors ($t<500$~ns) when within the mirrors' bandwidth. We also fix the steady-state populations by measuring the qubit population without any microwave excitation. This fit determines $\kappa_{ge}$.

We repeat the same single decaying exponential fit for the decay from $|f\rangle$ after excitation to $|f\rangle$, determining $\kappa_{ef}$. The $|e\rangle$ population evolution after excitation to $|f\rangle$ is modeled by Eq.~\ref{T1Formula} using the two fitted rates. The resulting fits are shown in Fig.~S3 for the operating frequencies $\omega_A $ and $\omega_B$ defined in the main text for $Q_1$, and agree very well with the data. The frequency dependence of $\kappa_{ge}$ and $\kappa_{ef}$ is shown in Fig.~2 of the main text.

We now consider the possibility of a  two-phonon relaxation process, under the hypothesis that it could  be strongly enhanced due to the frequency dependence of the IDT \cite{friskkockumDesigningFrequencydependentRelaxation2014}. The two-phonon relaxation is expected to be maximal when $\omega_{ge}-\vert\alpha\vert/2$ matches the IDT central frequency. We fit the population evolution from $|f\rangle$ using a two-parameter fit and keeping $\kappa_{ge}$ as given by fitting the decay from $|e\rangle$ after excitation to $|e\rangle$. The result is shown in Fig.~S3c. The extracted $\kappa_{gf}$ reaches a maximum of $1/30$~ns. At the operating frequencies of the main text, $\omega_A$ and $\omega_B$, the ratio $\kappa_{gf}/\kappa_{ef}$ is below $10\%$. The uncertainty of this determination is also quite large when $\kappa_{ef}$ is large - more than 50\%  whenever $\kappa_{ef}/2\pi \leqslant 20$~MHz. We thus conclude that even if this process occurs, it is negligible in our experiment.

\begin{figure}[tb]
\centering
\includegraphics[width=\textwidth]{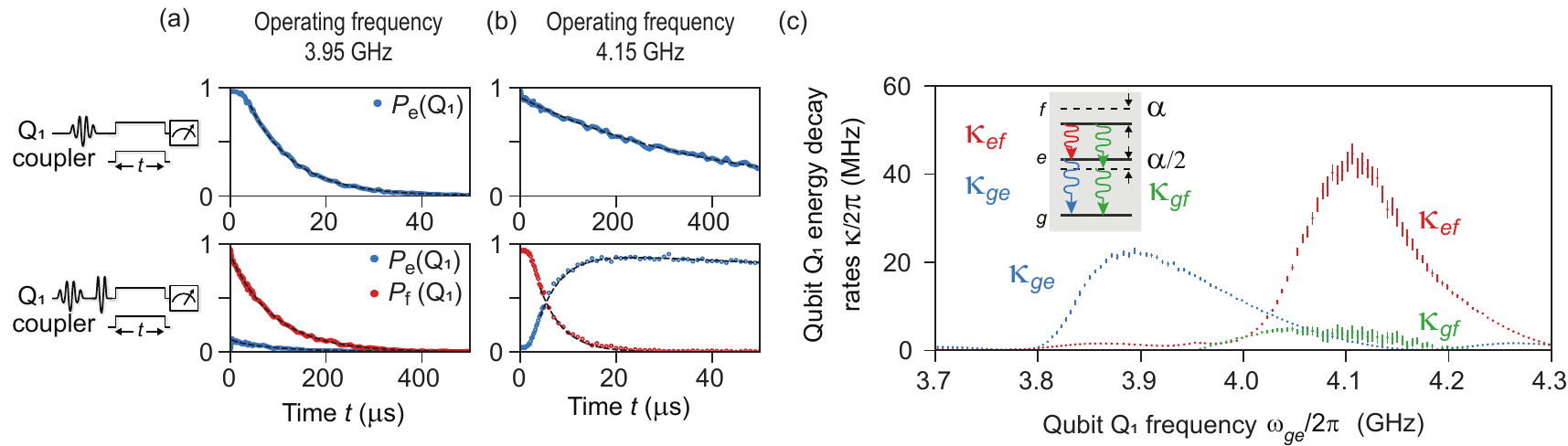}
\caption{\label{decay} (a-b) $Q_1$ energy decay rates are extracted for its first two transitions,  $\kappa_{ge}$ and $\kappa_{ef}$ at two operating points, $\omega_{ge}/2\pi = 3.95$~GHz (panel a) and $\omega_{ge}/2\pi=4.15$~GHz (panel b). The decay rates are fit from the exponential decay of $Q_1$'s state populations after excitation to $|e\rangle$ ($|f\rangle$), see top (bottom) panels. Dashed lines are the fits described in the text. (c) Extracted $Q_1$ energy decay rates when including a direct two-phonon relaxation from $|f\rangle$ to $|g\rangle$ with rate $\kappa_{gf}$.}
\centering
\end{figure}

We attempted to model the frequency-dependent relaxation rates using a circuit model for the qubit-coupler-IDT system. The IDT is modeled using a coupling-of-modes model \cite{morgandavidSurfaceAcousticWave2007}, taking into account the internal reflections occurring between the electrodes of the IDT, thus allowing us to infer the IDT admittance as a function of frequency, see Fig.~S4a. For reference, we also study the ideal case of an uniform transducer with no internal reflections, where the IDT admittance is given by
\begin{equation}
    Y_a(\omega) = i C_0 \omega + G_a(\omega) + i B_a(\omega), \label{simpleIDT}
\end{equation}
where $C_0$ is the IDT electrical capacitance, $G_a(\omega)$ the IDT conductance given in the main text, and $B_a$ the IDT susceptance related to its conductance by an Hilbert transformation $B_a(\omega) = G_a(\omega) \ast [-1/\pi\omega] $. We then derive the equivalent impedance $Z(\omega)$ for the circuit shown in Fig.~S1 looking into terminals A-B for qubit $Q_1$ (or C-D for qubit $Q_2$). We extract the circuit resonant frequencies by looking for the zeros of $Z(\omega)$ in the complex plane. Ignoring losses in the IDT (by setting $ \mathrm{Re}[Y(\omega)]=0$), we identify three modes: the qubit, the IDT series resonance, and the mode created by the IDT capacitor and the couplers' inductance networks. We then re-evaluate the frequencies of these modes in the presence of IDT loss. To extract the qubit relaxation rate and its anharmonicity, we approximate the circuit near the qubit resonance $\omega_q$ as an $RLC$ series circuit, with effective parameters
\begin{eqnarray}
L_{\mathrm{eff}} &=& 2/ \mathrm{Im}[Z'(\omega_q)],\\
C_{\mathrm{eff}} &=& 1/(L_{\mathrm{eff}}\omega_q^2),\\
R_{\mathrm{eff}} &=&  \mathrm{Re}[Z(\omega_q)].
\end{eqnarray}

The qubit relaxation rate $\kappa_{ge} = 1/T_1$ is then given by $T_1 \omega_q = Q$, where the qubit quality factor is given by $Q = \frac{\sqrt{L_{\mathrm{eff}}/C_{\mathrm{eff}}}}{R_{\mathrm{eff}}}$. To evaluate the anharmonicity, we split the effective qubit inductance $L_{\mathrm{eff}}$ into its non-linear part, arising from the qubit SQUID inductance $L_q(\phi_q)$, and its linear part $L_{\mathrm{eff}}-L_q(\phi_q)$. The anharmonicity is then given by \cite{chenQubitArchitectureHigh2014}:
\begin{equation}
\alpha = -\frac{e^2}{2C_{\mathrm{eff}}} \left(\frac{L_q(\phi_q)}{L_{\mathrm{eff}}}\right)^3.
\end{equation}
Finally, the relaxation rate from state $|f\rangle$ is given by $\kappa_{ef} = 2/T_{1,ef}$ where $T_{1,\mathrm{ef}} \times (\omega_q+\alpha) = Q_{\mathrm{ef}}$ and $Q_{\mathrm{ef}}$ is evaluated considering the following updated circuit parameters:
\begin{eqnarray}
L_{\mathrm{eff,ef}} &=& 2/ \mathrm{Im}[Z'(\omega_q+\alpha)],\\
C_{\mathrm{eff,ef}} &=& 1/[L_{\mathrm{eff,ef}}(\omega_q+\alpha)^2],\\
R_{\mathrm{eff,ef}} &=& \mathrm{Re}[Z(\omega_q+\alpha)],\\
Q_{\mathrm{ef}}&=& \frac{\sqrt{L_{\mathrm{eff,ef}}/C_{\mathrm{eff,ef}}}}{R_{\mathrm{eff,ef}}}.
\end{eqnarray}

To obtain the model relaxation rates shown in Fig.~2 of the main text, we use the parameters listed in Table S1 as input parameters. The non-design parameters were calibrated as follow: The qubit capacitance was adjusted to reproduce the qubit anharmonicity and the qubit SQUID inductance was adjusted to reproduce the measured qubit bare frequency. The couplers' Josephson junction inductances were calibrated using the qubit frequency shift induced by the coupler, with the qubit tuned at a non-zero emission point for the IDT. The mutual inductive coupling between the two chips was calibrated using the qubit-qubit direct electrical coupling ($\sim g/2\pi = 1.1~$MHz) at a non-zero emission point for the IDT. The SAW velocity for the IDT was adjusted to match the frequencies of the two zero emission points. Finally the IDT reflectivity $r$ is an imaginary free parameter, whose value is expected to be small ($|r|\lesssim 1$\%) for a 30-nm-thick aluminum transducer fabricated on a $128^\circ Y-X$ lithium niobate wafer \cite{morgandavidSurfaceAcousticWave2007}.

The IDT internal reflectivity ($r = 0.009i$) and the mutual inductance ($M = 0.23$~pH) between the chips are adjusted to match as closely as possible the $\kappa_{eg}$ curve. As can be seen in Fig.~2 of the main text, this model can reproduce qualitatively the  $\kappa_{eg}$ rates, but only roughly matches the $\kappa_{ef}$ measured rates, with a significant 50-MHz discrepancy for the $\kappa_{ef}$ maximum. Tuning the parameters of the IDT (capacitance and reflectivity) and of the coupling circuit (mutual inductance between the flip-chips and the couplers' junction inductances) does not give a better agreement.

\begin{figure}[tb]
\centering
\includegraphics[width=\textwidth]{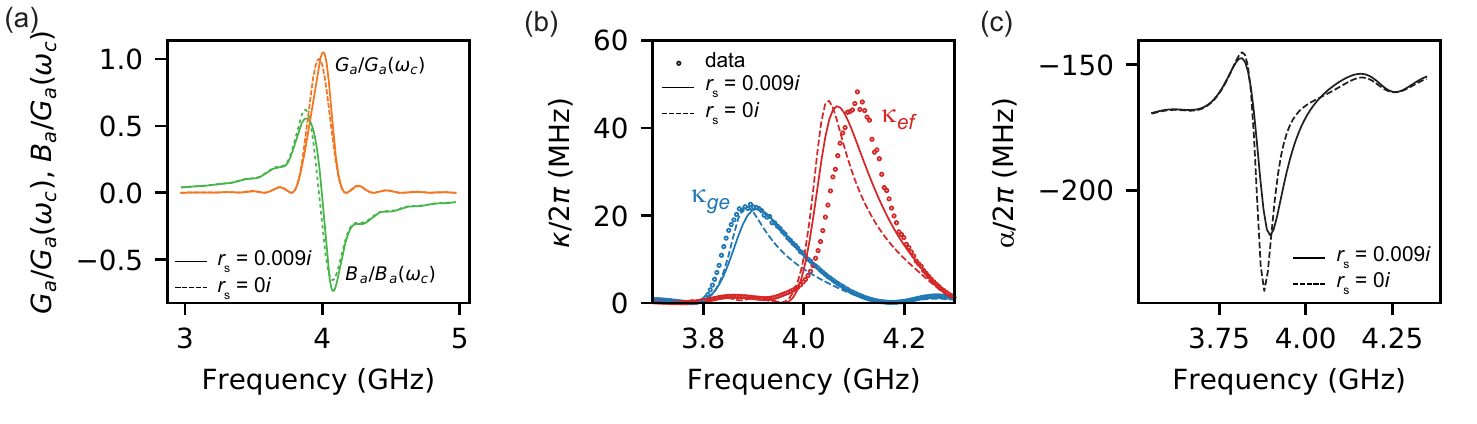}
\caption{\label{decay2} (a) Conductance (orange) and susceptance (green) for a $20$ finger pairs uniform IDT, with no internal reflectivity (dashed lines) and a small reflectivity (solid lines). (b) Energy decay rates $\kappa_{ge}$ (blue) and $\kappa_{ef}$ (red)  extracted for the electrical circuit considered in Fig.~S1, as detailed in the text, considering a non-reflective transducer (dashed lines, $r=0$, $M = 0.23$~pH) and a small amount of internal reflections (solid lines, $r = 0.009j$, $M=0.21$~pH). The mutual coupling between the two chips was adjusted to reproduce the height of the $\kappa_{g\-e}$ peak. (c) Induced qubit anharmonicity for both types of transducers. The error in $\kappa_{e\-f}$ seems to arise from an overestimate of the anharmonicity.}
\centering
\end{figure}

By comparing the model derived using the simple symmetric IDT admittance given by Eq.~\ref{simpleIDT}, see Fig.~\ref{decay2}, we see that most of the asymmetry of the $\kappa_{eg}$ curve is actually due to the coupling of the qubit to the IDT, and not to the internal reflections of the IDT. This can be understood as the Lamb shift induced on the qubit \cite{friskkockumDesigningFrequencydependentRelaxation2014} when coupled to the IDT, which is related to $B_a(\omega)$ \cite{friskkockumDesigningFrequencydependentRelaxation2014} and which also induces a frequency dependence for the qubit anharmonicity when coupled to the IDT.

\section{Numerical modeling}

In this last section, we address the numerical modeling of the quantum eraser experiment, as well as the transfers from Fig.~3 in the main text. In the quantum eraser experiment, the system comprises two qubits ($Q_1$ and $Q_2$ and the itinerant wavepackets corresponding to the phonons $A$ and $B$.

We model the qubits as anharmonic oscillators with bosonic creation operators $\hat{s}_i$. Their non-interacting Hamiltonians in the frame rotating at the frequency of phonon $A$ is given by 
\begin{equation}
    H_{0,i}/\hbar = \Delta_i \hat{s}_i^{\dagger} \hat{s}_i + \alpha_i  \hat{s}_i^{\dagger} \hat{s}_i^{\dagger} \hat{s}_i \hat{s}_i,
\end{equation}
where $\Delta_i$ is the detuning of qubit $i$ with respect to phonon $A$ and $\alpha_i$ its anharmonicity. We also define the qubit matrix element operators $\hat{s}_{\mathrm{ge},i} = |g\rangle\langle e|$ and $\hat{s}_{\mathrm{ef},i} = |e\rangle\langle f|$ to take into account transition-dependent effects.

The itinerant wavepackets are modeled as bosonic modes. To accurately describe the emission and capture of the phonons, and model the evolution of the populations in these itinerant bosonic modes, we use the theory derived in \cite{kiilerichInputOutputTheoryQuantum2019}. Each interaction of the qubits with the acoustic channel requires the use of two wavepackets: an input wavepacket and an output wavepacket. As we wish to model four interactions (the half-emission (or the herald emission) and the half-phonon capture (or the herald capture) for phonon mode A (or B)), we only need to consider six wavepackets: First, $a_{\mathrm{in}}(t)$ ($b_{\mathrm{in}}(t)$), the input acoustic field that interacts with qubit $Q_1$ during the phonon A (B) emission at time $t_{e,A}$ ($t_{e,B}$). Second, $a_{\mathrm{rt}}(t)$ ($b_{\mathrm{rt}}(t)$) is the acoustic output field describing the result from the interaction of $a_{\mathrm{in}}(t)$ ($b_{\mathrm{in}}(t)$) with the qubit and contains the emitted phonon A (B). This field will then be used as input for the second interaction with the qubit after it completes one round-trip during the capture process. Third, the field $c_{\mathrm{out}}(t)$ ($b_{\mathrm{out}}(t)$), the acoustic output field containing the phonon resulting from this second interaction. According to \cite{kiilerichInputOutputTheoryQuantum2019}, bosonic annihilation operators can be used to describe the quantum state contained in these wavepackets, defined as
\begin{eqnarray}
\hat{w} &=& \int w(t) u_{w}(t) dt,
\end{eqnarray}
where the functions $u_w(t)$ describe the wavepacket envelopes and are normalized such that $\int{|u_w(t)|^2 dt} = 1$.

To describe the interactions of the qubits with the acoustic channel during the quantum eraser experiment, we thus use six bosonic creation operators, three ($\hat{a}_{\mathrm{in}}$, $\hat{a}_{\mathrm{rt}}$, and $\hat{a}_{\mathrm{out}}$) for the half-phonon and three ($\hat{b}_{\mathrm{in}}$, $\hat{b}_{\mathrm{rt}}$, and $\hat{b}_{\mathrm{out}}$) for the heralding phonon. We note that $\hat{a}_{\mathrm{rt}}$ and $\hat{b}_{\mathrm{rt}}$ correspond to what we call phonons $A$ and $B$ in the main text.

In the frame rotating at the phonon $A$ emission frequency, the stationary Hamiltonian of the system is
\begin{equation}
H_0/\hbar = \sum_{i=1,2} H_{0,i}/\hbar + \sum_{i=\mathrm{in},\mathrm{rt},\mathrm{out}}\Delta_{b,i} \hat{b}_i^{\dagger} \hat{b}_i
\end{equation}
where $\Delta_{b,i}$ is the detuning of phonon $B$ with respect to the frequency of phonon $A$.

The interaction of the sub-system comprising one incoming bosonic mode $\hat{a}_x$ and one outgoing bosonic mode $\hat{a}_y$ interacting with one of the qubits at time $t$, either on its $g \textrm{-} e$ or $e \textrm{-} f$ transition, $\hat{c} = \hat{s}_{\mathrm{ge},i}$ or $\hat{c}= \hat{s}_{\mathrm{ef},i}$, at the coupling rate $\kappa_i(t)$ set by the coupler, is described by this master equation:
\begin{equation}\label{meq}
\dot{\rho} (t) = - \frac{i}{\hbar} [\rho(t), \hat{H}(t)] + \hat{L}_0\rho\hat{L}_0^{\dagger} - \frac{1}{2} \lbrace \hat{L}_0^{\dagger} \hat{L}_0, \rho(t) \rbrace,
\end{equation}
where the Hamiltonian $\hat{H}(t)$ is given by
\begin{equation}
\hat{H}/\hbar = \hat{H_0}/\hbar + \frac{i}{2}\left(\sqrt{\kappa_i(t)} g_{\mathrm{in}}^\ast(t) \hat{a}_{x}^{\dagger} \hat{c} + \sqrt{\kappa_i(t)} g_{\mathrm{out}}(t) \hat{c}^{\dagger} \hat{a}_{y} + g_{\mathrm{in}}^\ast(t) g_{\mathrm{out}}(t) \hat{a}_{x}^{\dagger} \hat{a}_{y} - h.c. \right),
\end{equation}
and the Lindblad operator is given by
\begin{equation}
\hat{L}_0(t) = \sqrt{\kappa_i(t)} \hat{c} + g_{\mathrm{in}}(t) \hat{a}_{x} + g_{\mathrm{out}}(t) \hat{a}_{y}.
\end{equation}
In the above equations, the coupling coefficients are given by
\begin{eqnarray}
g_{\mathrm{in}}(t') &=& \frac{\sqrt{\kappa_c}}{\sqrt{1+e^{-\kappa_c t'}}}, ~\textrm{and} \\
g_{\mathrm{out}}(t') &=& -\frac{\sqrt{\kappa_c}}{\sqrt{1+e^{\kappa_c t'}}},
\end{eqnarray}
using the cosecant wavepackets from the experiment.

We simulate the total evolution of the system using four consecutive integrations of Eq.~\ref{meq}. In addition, we include the action of qubit decoherence and acoustic losses by including the following Lindblad dissipation operators: $\sqrt{1/T_1} \hat{s}_i$ for the intrinsic qubit relaxation, $\sqrt{1/T_{\phi}} \hat{s}_{\mathrm{ge}}^{\dagger}\hat{s}_{\mathrm{ge}}$ and $\sqrt{1/T_{\phi}} \hat{s}_{\mathrm{ef}}^{\dagger}\hat{s}_{\mathrm{ef}}$ for the qubit decoherence for the $g \textrm{-} e$ and $e \textrm{-} f$ transitions, with $1/T_{\phi,\mathrm{ge}|\mathrm{ef}} = 1/T_{2,R,\mathrm{ge}|\mathrm{ef}} - 1/(2T_1)$, and $\sqrt{\kappa_a} \hat{a}$, $\sqrt{\kappa_b} \hat{b}$ for the acoustic losses, with $\kappa_a$ and $\kappa_b$ defined to match the round-trip transfer efficiency $\eta_a = e^{-\kappa_a\tau}$ and $\eta_b = e^{-\kappa_b\tau}$. The qubits' $XY$ drives are modeled using $H_D/\hbar = \beta (\hat{s}_i e^{i \omega_d t}+\hat{s}_i^{\dagger} e^{-i \omega_d t}) $, where $\beta$ is adjusted to give the measured rotation.

We perform these master equations simulations using QuTip \cite{johanssonQuTiPOpensourcePython2012}, with the control sequences defined in Fig.~3 and 4 of the main text as inputs. The model input parameters are given in Table~S1. The extracted populations are shown in Fig.~3 and 4 of the main text, corrected for readout errors only in Fig.~4b, and giving good agreement with the measured data.

\renewcommand{\baselinestretch}{1.}
\begin{table}[h]
\center
  \begin{tabular}{p{9cm}ccc}
    \hline
    \hline
    Qubit parameters & \  & Qubit 1 & Qubit 2\\
    \hline
    Qubit bare frequency (GHz) & \  & 4.86 & $\sim 6$\\
    Qubit capacitance (fF) & \  & 100 & 100\\
    SQUID inductance (nH) & \  & 10.4 & 7.0\\
    Qubit anharmonicity (MHz) & \  & -179 & -188\\
    Qubit intrinsic lifetime, eg transition, $T_{1, eg, \mathrm{int}}$ ($\mu$s) & \
    & 18 & 18\\
    Qubit intrinsic lifetime, ef transition, $T_{1, ef, \mathrm{int}}$ ($\mu$s) & \
    & 11 & 11\\
    Qubit Ramsey dephasing time,  eg transition, $T_{2, ge,\mathrm{Ramsey}}$ ($\mu$s) & \  &
    1.2 & 0.8\\
    Qubit Ramsey dephasing time, ef transition, $T_{2, ef,\mathrm{Ramsey}}$ ($\mu$s) & \  &
    0.4 & 0.4\\
    $| f \rangle$ state readout fidelity & \  & 0.93 & 0.92\\
    $| e \rangle$ state readout fidelity & \  & 0.97 & 0.95\\
    $| g \rangle$ state readout fidelity & \  & 0.99 & 0.99\\
    \hline \hline
    \  & \  & \  & \ \\
    \hline \hline
    Tunable coupler parameters & \  & Coupler 1 & Coupler 2\\
    \hline
    Coupler junction inductance (nH) & \  & 1.19 & 1.24\\
    IDT grounding inductance (design value) (nH) & \  & 0.4 & 0.4\\
    Coupler grounding inductance (design value) (nH) & \  & 0.4 & 0.4\\
    Mutual coupling inductance between IDT and coupler (nH) & \  & 0.21 &
    0.21\\
    \hline \hline
    \  & \  & \  & \ \\
    \hline \hline
    SAW resonator parameters & Free space & Mirror  & Transducer\\
    \hline
    Aperture (\textmu m) & \  & 75 & 75\\
    Wave propagation speed (km/s) & 4.034(2) & 3.928(2) & 3.911(2)\\
    Wave propagation losses (Np/m) & 70(10) & - & -\\
    Number of cells & \  & 400 & 20\\
    Pitch (\textmu m) & \  & 0.5 & 0.985\\
    Reflectivity & \  & -0.049$i$(5) & 0.009$i$(2)\\
    Metallization ratio & \  & 0.58 & 0.58\\
    Effective mirror-mirror distance ($\mu$m) &   & 2029.6 &   \\
    Free spectral range (MHz) &  & 1.97 &   \\
    \hline \hline
  \end{tabular}
  \caption{\label{tableone}Device parameters for the two qubits, parameters related to the interdigitated acoustic transducer (IDT), the tunable couplers connecting each qubit to the SAW resonator, and the SAW resonator itself.}
\end{table}
\renewcommand{\baselinestretch}{1.5}

\clearpage

merlin.mbs apsrev4-1.bst 2010-07-25 4.21a (PWD, AO, DPC) hacked


\begin{thebibliography}{48}%
\makeatletter
\providecommand \@ifxundefined [1]{%
 \@ifx{#1\undefined}
}%
\providecommand \@ifnum [1]{%
 \ifnum #1\expandafter \@firstoftwo
 \else \expandafter \@secondoftwo
 \fi
}%
\providecommand \@ifx [1]{%
 \ifx #1\expandafter \@firstoftwo
 \else \expandafter \@secondoftwo
 \fi
}%
\providecommand \natexlab [1]{#1}%
\providecommand \enquote  [1]{``#1''}%
\providecommand \bibnamefont  [1]{#1}%
\providecommand \bibfnamefont [1]{#1}%
\providecommand \citenamefont [1]{#1}%
\providecommand \href@noop [0]{\@secondoftwo}%
\providecommand \href [0]{\begingroup \@sanitize@url \@href}%
\providecommand \@href[1]{\@@startlink{#1}\@@href}%
\providecommand \@@href[1]{\endgroup#1\@@endlink}%
\providecommand \@sanitize@url [0]{\catcode `\\12\catcode `\$12\catcode
  `\&12\catcode `\#12\catcode `\^12\catcode `\_12\catcode `\%12\relax}%
\providecommand \@@startlink[1]{}%
\providecommand \@@endlink[0]{}%
\providecommand \url  [0]{\begingroup\@sanitize@url \@url }%
\providecommand \@url [1]{\endgroup\@href {#1}{\urlprefix }}%
\providecommand \urlprefix  [0]{URL }%
\providecommand \Eprint [0]{\href }%
\providecommand \doibase [0]{http://dx.doi.org/}%
\providecommand \selectlanguage [0]{\@gobble}%
\providecommand \bibinfo  [0]{\@secondoftwo}%
\providecommand \bibfield  [0]{\@secondoftwo}%
\providecommand \translation [1]{[#1]}%
\providecommand \BibitemOpen [0]{}%
\providecommand \bibitemStop [0]{}%
\providecommand \bibitemNoStop [0]{.\EOS\space}%
\providecommand \EOS [0]{\spacefactor3000\relax}%
\providecommand \BibitemShut  [1]{\csname bibitem#1\endcsname}%
\let\auto@bib@innerbib\@empty
%</preamble>
\bibitem [{\citenamefont {Young}(1807)}]{youngCourseLecturesNatural1807}%
  \BibitemOpen
  \bibfield  {author} {\bibinfo {author} {\bibfnamefont {Thomas}\ \bibnamefont
  {Young}},\ }\href@noop {} {\emph {\bibinfo {title} {A Course of Lectures on
  Natural Philosophy and the Mechanical Arts.}}}\ (\bibinfo  {publisher}
  {{Printed for J. Johnson,}},\ \bibinfo {address} {{London :}},\ \bibinfo
  {year} {1807})\BibitemShut {NoStop}%
\bibitem [{\citenamefont {Bohr}(1949)}]{bohrdiscussion}%
  \BibitemOpen
  \bibfield  {author} {\bibinfo {author} {\bibfnamefont {Niels}\ \bibnamefont
  {Bohr}},\ }in\ \href@noop {} {\emph {\bibinfo {booktitle} {Albert Einstein:
  Philosopher Scientist}}},\ \bibinfo {editor} {edited by\ \bibinfo {editor}
  {\bibfnamefont {P.~A.}\ \bibnamefont {Schilpp}}}\ (\bibinfo  {publisher}
  {Library of Living Philosophers, Evanston},\ \bibinfo {year} {1949})\ pp.\
  \bibinfo {pages} {200--241},\ \bibinfo {note} {reprinted in Quantum Theory
  and Measurement (eds Wheeler, J. A. and Zurek, W. H.) 9--49 (Princeton Univ.
  Press, Princeton, 1983)}\BibitemShut {NoStop}%
\bibitem [{\citenamefont {Wheeler}(1978)}]{wheeler1978}%
  \BibitemOpen
  \bibfield  {author} {\bibinfo {author} {\bibfnamefont {J.~A.}\ \bibnamefont
  {Wheeler}},\ }\bibfield  {title} {\enquote {\bibinfo {title} {The past and
  the delayed-choice double-slit experiment},}\ }in\ \href@noop {} {\emph
  {\bibinfo {booktitle} {Mathematical Foundations of Quantum Theory}}},\
  \bibinfo {editor} {edited by\ \bibinfo {editor} {\bibfnamefont {A.~R.}\
  \bibnamefont {Marlow}}}\ (\bibinfo  {publisher} {Academic Press},\ \bibinfo
  {year} {1978})\ pp.\ \bibinfo {pages} {9--48}\BibitemShut {NoStop}%
\bibitem [{\citenamefont {Scully}\ and\ \citenamefont
  {Dr{\"u}hl}(1982)}]{scullyQuantumEraserProposed1982}%
  \BibitemOpen
  \bibfield  {author} {\bibinfo {author} {\bibfnamefont {Marlan~O.}\
  \bibnamefont {Scully}}\ and\ \bibinfo {author} {\bibfnamefont {Kai}\
  \bibnamefont {Dr{\"u}hl}},\ }\bibfield  {title} {\enquote {\bibinfo {title}
  {Quantum eraser: {{A}} proposed photon correlation experiment concerning
  observation and "delayed choice" in quantum mechanics},}\ }\href {\doibase
  10.1103/PhysRevA.25.2208} {\bibfield  {journal} {\bibinfo  {journal}
  {Physical Review A}\ }\textbf {\bibinfo {volume} {25}},\ \bibinfo {pages}
  {2208--2213} (\bibinfo {year} {1982})}\BibitemShut {NoStop}%
\bibitem [{\citenamefont {Wheeler}(1984)}]{wheeler1984}%
  \BibitemOpen
  \bibfield  {author} {\bibinfo {author} {\bibfnamefont {J.~A.}\ \bibnamefont
  {Wheeler}},\ }\bibfield  {title} {\enquote {\bibinfo {title} {Law without
  law},}\ }in\ \href@noop {} {\emph {\bibinfo {booktitle} {Quantum Theory and
  Measurement}}},\ \bibinfo {editor} {edited by\ \bibinfo {editor}
  {\bibfnamefont {J.~A.}\ \bibnamefont {Wheeler}}\ and\ \bibinfo {editor}
  {\bibfnamefont {W.~H.}\ \bibnamefont {Zurek}}}\ (\bibinfo  {publisher}
  {Princeton University Press},\ \bibinfo {year} {1984})\ pp.\ \bibinfo {pages}
  {182--213}\BibitemShut {NoStop}%
\bibitem [{\citenamefont {Kwiat}\ \emph {et~al.}(1992)\citenamefont {Kwiat},
  \citenamefont {Steinberg},\ and\ \citenamefont
  {Chiao}}]{kwiatObservationQuantumEraser1992}%
  \BibitemOpen
  \bibfield  {author} {\bibinfo {author} {\bibfnamefont {Paul~G.}\ \bibnamefont
  {Kwiat}}, \bibinfo {author} {\bibfnamefont {Aephraim~M.}\ \bibnamefont
  {Steinberg}}, \ and\ \bibinfo {author} {\bibfnamefont {Raymond~Y.}\
  \bibnamefont {Chiao}},\ }\bibfield  {title} {\enquote {\bibinfo {title}
  {Observation of a ``quantum eraser'': {{A}} revival of coherence in a
  two-photon interference experiment},}\ }\href {\doibase
  10.1103/PhysRevA.45.7729} {\bibfield  {journal} {\bibinfo  {journal}
  {Physical Review A}\ }\textbf {\bibinfo {volume} {45}},\ \bibinfo {pages}
  {7729--7739} (\bibinfo {year} {1992})}\BibitemShut {NoStop}%
\bibitem [{\citenamefont {Kim}\ \emph {et~al.}(2000)\citenamefont {Kim},
  \citenamefont {Yu}, \citenamefont {Kulik}, \citenamefont {Shih},\ and\
  \citenamefont {Scully}}]{kimDelayedChoiceQuantum2000}%
  \BibitemOpen
  \bibfield  {author} {\bibinfo {author} {\bibfnamefont {Yoon-Ho}\ \bibnamefont
  {Kim}}, \bibinfo {author} {\bibfnamefont {Rong}\ \bibnamefont {Yu}}, \bibinfo
  {author} {\bibfnamefont {Sergei~P.}\ \bibnamefont {Kulik}}, \bibinfo {author}
  {\bibfnamefont {Yanhua}\ \bibnamefont {Shih}}, \ and\ \bibinfo {author}
  {\bibfnamefont {Marlan~O.}\ \bibnamefont {Scully}},\ }\bibfield  {title}
  {\enquote {\bibinfo {title} {Delayed ``{{Choice}}'' {{Quantum Eraser}}},}\
  }\href {\doibase 10.1103/PhysRevLett.84.1} {\bibfield  {journal} {\bibinfo
  {journal} {Physical Review Letters}\ }\textbf {\bibinfo {volume} {84}},\
  \bibinfo {pages} {1--5} (\bibinfo {year} {2000})}\BibitemShut {NoStop}%
\bibitem [{\citenamefont {Ma}\ \emph {et~al.}(2013)\citenamefont {Ma},
  \citenamefont {Kofler}, \citenamefont {Qarry}, \citenamefont {Tetik},
  \citenamefont {Scheidl}, \citenamefont {Ursin}, \citenamefont {Ramelow},
  \citenamefont {Herbst}, \citenamefont {Ratschbacher}, \citenamefont
  {Fedrizzi}, \citenamefont {Jennewein},\ and\ \citenamefont
  {Zeilinger}}]{maQuantumErasureCausally2013}%
  \BibitemOpen
  \bibfield  {author} {\bibinfo {author} {\bibfnamefont {Xiao-Song}\
  \bibnamefont {Ma}}, \bibinfo {author} {\bibfnamefont {Johannes}\ \bibnamefont
  {Kofler}}, \bibinfo {author} {\bibfnamefont {Angie}\ \bibnamefont {Qarry}},
  \bibinfo {author} {\bibfnamefont {Nuray}\ \bibnamefont {Tetik}}, \bibinfo
  {author} {\bibfnamefont {Thomas}\ \bibnamefont {Scheidl}}, \bibinfo {author}
  {\bibfnamefont {Rupert}\ \bibnamefont {Ursin}}, \bibinfo {author}
  {\bibfnamefont {Sven}\ \bibnamefont {Ramelow}}, \bibinfo {author}
  {\bibfnamefont {Thomas}\ \bibnamefont {Herbst}}, \bibinfo {author}
  {\bibfnamefont {Lothar}\ \bibnamefont {Ratschbacher}}, \bibinfo {author}
  {\bibfnamefont {Alessandro}\ \bibnamefont {Fedrizzi}}, \bibinfo {author}
  {\bibfnamefont {Thomas}\ \bibnamefont {Jennewein}}, \ and\ \bibinfo {author}
  {\bibfnamefont {Anton}\ \bibnamefont {Zeilinger}},\ }\bibfield  {title}
  {\enquote {\bibinfo {title} {Quantum erasure with causally disconnected
  choice},}\ }\href {\doibase 10.1073/pnas.1213201110} {\bibfield  {journal}
  {\bibinfo  {journal} {Proceedings of the National Academy of Sciences}\
  }\textbf {\bibinfo {volume} {110}},\ \bibinfo {pages} {1221--1226} (\bibinfo
  {year} {2013})}\BibitemShut {NoStop}%
\bibitem [{\citenamefont {Liu}\ \emph {et~al.}(2017)\citenamefont {Liu},
  \citenamefont {Xu}, \citenamefont {Wang}, \citenamefont {Zheng},
  \citenamefont {Roy}, \citenamefont {Kundu}, \citenamefont {Chand},
  \citenamefont {Ranadive}, \citenamefont {Vijay}, \citenamefont {Song},
  \citenamefont {Duan},\ and\ \citenamefont
  {Sun}}]{liuTwofoldQuantumDelayedchoice2017}%
  \BibitemOpen
  \bibfield  {author} {\bibinfo {author} {\bibfnamefont {Ke}~\bibnamefont
  {Liu}}, \bibinfo {author} {\bibfnamefont {Yuan}\ \bibnamefont {Xu}}, \bibinfo
  {author} {\bibfnamefont {Weiting}\ \bibnamefont {Wang}}, \bibinfo {author}
  {\bibfnamefont {Shi-Biao}\ \bibnamefont {Zheng}}, \bibinfo {author}
  {\bibfnamefont {Tanay}\ \bibnamefont {Roy}}, \bibinfo {author} {\bibfnamefont
  {Suman}\ \bibnamefont {Kundu}}, \bibinfo {author} {\bibfnamefont {Madhavi}\
  \bibnamefont {Chand}}, \bibinfo {author} {\bibfnamefont {Arpit}\ \bibnamefont
  {Ranadive}}, \bibinfo {author} {\bibfnamefont {Rajamani}\ \bibnamefont
  {Vijay}}, \bibinfo {author} {\bibfnamefont {Yipu}\ \bibnamefont {Song}},
  \bibinfo {author} {\bibfnamefont {Luming}\ \bibnamefont {Duan}}, \ and\
  \bibinfo {author} {\bibfnamefont {Luyan}\ \bibnamefont {Sun}},\ }\bibfield
  {title} {\enquote {\bibinfo {title} {A twofold quantum delayed-choice
  experiment in a superconducting circuit},}\ }\href {\doibase
  10.1126/sciadv.1603159} {\bibfield  {journal} {\bibinfo  {journal} {Science
  Advances}\ }\textbf {\bibinfo {volume} {3}},\ \bibinfo {pages} {e1603159}
  (\bibinfo {year} {2017})}\BibitemShut {NoStop}%
\bibitem [{\citenamefont {Morgan}(2007)}]{morgandavidSurfaceAcousticWave2007}%
  \BibitemOpen
  \bibfield  {author} {\bibinfo {author} {\bibfnamefont {David}\ \bibnamefont
  {Morgan}},\ }\href {\doibase 10.1016/B978-0-12-372537-0.X5000-6} {\emph
  {\bibinfo {title} {Surface {{Acoustic Wave Filters}}}}}\ (\bibinfo
  {publisher} {{Elsevier}},\ \bibinfo {year} {2007})\BibitemShut {NoStop}%
\bibitem [{\citenamefont {Bienfait}\ \emph {et~al.}(2019)\citenamefont
  {Bienfait}, \citenamefont {Satzinger}, \citenamefont {Zhong}, \citenamefont
  {Chang}, \citenamefont {Chou}, \citenamefont {Conner}, \citenamefont {Dumur},
  \citenamefont {Grebel}, \citenamefont {Peairs}, \citenamefont {Povey},\ and\
  \citenamefont {Cleland}}]{bienfaitPhononmediatedQuantumState2019}%
  \BibitemOpen
  \bibfield  {author} {\bibinfo {author} {\bibfnamefont {A.}~\bibnamefont
  {Bienfait}}, \bibinfo {author} {\bibfnamefont {K.~J.}\ \bibnamefont
  {Satzinger}}, \bibinfo {author} {\bibfnamefont {Y.~P.}\ \bibnamefont
  {Zhong}}, \bibinfo {author} {\bibfnamefont {H.-S.}\ \bibnamefont {Chang}},
  \bibinfo {author} {\bibfnamefont {M.-H.}\ \bibnamefont {Chou}}, \bibinfo
  {author} {\bibfnamefont {C.~R.}\ \bibnamefont {Conner}}, \bibinfo {author}
  {\bibfnamefont {{\'E}}~\bibnamefont {Dumur}}, \bibinfo {author}
  {\bibfnamefont {J.}~\bibnamefont {Grebel}}, \bibinfo {author} {\bibfnamefont
  {G.~A.}\ \bibnamefont {Peairs}}, \bibinfo {author} {\bibfnamefont {R.~G.}\
  \bibnamefont {Povey}}, \ and\ \bibinfo {author} {\bibfnamefont {A.~N.}\
  \bibnamefont {Cleland}},\ }\bibfield  {title} {\enquote {\bibinfo {title}
  {Phonon-mediated quantum state transfer and remote qubit entanglement},}\
  }\href {\doibase 10.1126/science.aaw8415} {\bibfield  {journal} {\bibinfo
  {journal} {Science}\ }\textbf {\bibinfo {volume} {364}},\ \bibinfo {pages}
  {368--371} (\bibinfo {year} {2019})}\BibitemShut {NoStop}%
\bibitem [{\citenamefont {Delsing}\ \emph {et~al.}(2019)\citenamefont
  {Delsing}, \citenamefont {Cleland}, \citenamefont {Schuetz}, \citenamefont
  {Kn{\"o}rzer}, \citenamefont {Giedke}, \citenamefont {Cirac}, \citenamefont
  {Srinivasan}, \citenamefont {Wu}, \citenamefont {Balram}, \citenamefont
  {B{\"a}uerle}, \citenamefont {Meunier}, \citenamefont {Ford}, \citenamefont
  {Santos}, \citenamefont {{Cerda-M{\'e}ndez}}, \citenamefont {Wang},
  \citenamefont {Krenner}, \citenamefont {Nysten}, \citenamefont {{s}},
  \citenamefont {Nash}, \citenamefont {Thevenard}, \citenamefont {Gourdon},
  \citenamefont {Rovillain}, \citenamefont {Marangolo}, \citenamefont
  {Duquesne}, \citenamefont {Fischerauer}, \citenamefont {Ruile}, \citenamefont
  {Reiner}, \citenamefont {Paschke}, \citenamefont {Denysenko}, \citenamefont
  {Volkmer}, \citenamefont {Wixforth}, \citenamefont {Bruus}, \citenamefont
  {Wiklund}, \citenamefont {Reboud}, \citenamefont {Cooper}, \citenamefont
  {Fu}, \citenamefont {Brugger}, \citenamefont {Rehfeldt},\ and\ \citenamefont
  {Westerhausen}}]{delsing2019SurfaceAcoustic2019}%
  \BibitemOpen
  \bibfield  {author} {\bibinfo {author} {\bibfnamefont {Per}\ \bibnamefont
  {Delsing}}, \bibinfo {author} {\bibfnamefont {Andrew~N.}\ \bibnamefont
  {Cleland}}, \bibinfo {author} {\bibfnamefont {Martin J.~A.}\ \bibnamefont
  {Schuetz}}, \bibinfo {author} {\bibfnamefont {Johannes}\ \bibnamefont
  {Kn{\"o}rzer}}, \bibinfo {author} {\bibfnamefont {G{\'e}za}\ \bibnamefont
  {Giedke}}, \bibinfo {author} {\bibfnamefont {J.~Ignacio}\ \bibnamefont
  {Cirac}}, \bibinfo {author} {\bibfnamefont {Kartik}\ \bibnamefont
  {Srinivasan}}, \bibinfo {author} {\bibfnamefont {Marcelo}\ \bibnamefont
  {Wu}}, \bibinfo {author} {\bibfnamefont {Krishna~Coimbatore}\ \bibnamefont
  {Balram}}, \bibinfo {author} {\bibfnamefont {Christopher}\ \bibnamefont
  {B{\"a}uerle}}, \bibinfo {author} {\bibfnamefont {Tristan}\ \bibnamefont
  {Meunier}}, \bibinfo {author} {\bibfnamefont {Christopher J.~B.}\
  \bibnamefont {Ford}}, \bibinfo {author} {\bibfnamefont {Paulo~V.}\
  \bibnamefont {Santos}}, \bibinfo {author} {\bibfnamefont {Edgar}\
  \bibnamefont {{Cerda-M{\'e}ndez}}}, \bibinfo {author} {\bibfnamefont
  {Hailin}\ \bibnamefont {Wang}}, \bibinfo {author} {\bibfnamefont {Hubert~J.}\
  \bibnamefont {Krenner}}, \bibinfo {author} {\bibfnamefont {Emeline D.~S.}\
  \bibnamefont {Nysten}}, \bibinfo {author} {\bibfnamefont
  {Matthias~Wei\textbackslash{}s}\ \bibnamefont {{s}}}, \bibinfo {author}
  {\bibfnamefont {Geoff~R.}\ \bibnamefont {Nash}}, \bibinfo {author}
  {\bibfnamefont {Laura}\ \bibnamefont {Thevenard}}, \bibinfo {author}
  {\bibfnamefont {Catherine}\ \bibnamefont {Gourdon}}, \bibinfo {author}
  {\bibfnamefont {Pauline}\ \bibnamefont {Rovillain}}, \bibinfo {author}
  {\bibfnamefont {Max}\ \bibnamefont {Marangolo}}, \bibinfo {author}
  {\bibfnamefont {Jean-Yves}\ \bibnamefont {Duquesne}}, \bibinfo {author}
  {\bibfnamefont {Gerhard}\ \bibnamefont {Fischerauer}}, \bibinfo {author}
  {\bibfnamefont {Werner}\ \bibnamefont {Ruile}}, \bibinfo {author}
  {\bibfnamefont {Alexander}\ \bibnamefont {Reiner}}, \bibinfo {author}
  {\bibfnamefont {Ben}\ \bibnamefont {Paschke}}, \bibinfo {author}
  {\bibfnamefont {Dmytro}\ \bibnamefont {Denysenko}}, \bibinfo {author}
  {\bibfnamefont {Dirk}\ \bibnamefont {Volkmer}}, \bibinfo {author}
  {\bibfnamefont {Achim}\ \bibnamefont {Wixforth}}, \bibinfo {author}
  {\bibfnamefont {Henrik}\ \bibnamefont {Bruus}}, \bibinfo {author}
  {\bibfnamefont {Martin}\ \bibnamefont {Wiklund}}, \bibinfo {author}
  {\bibfnamefont {Julien}\ \bibnamefont {Reboud}}, \bibinfo {author}
  {\bibfnamefont {Jonathan~M.}\ \bibnamefont {Cooper}}, \bibinfo {author}
  {\bibfnamefont {YongQing}\ \bibnamefont {Fu}}, \bibinfo {author}
  {\bibfnamefont {Manuel~S.}\ \bibnamefont {Brugger}}, \bibinfo {author}
  {\bibfnamefont {Florian}\ \bibnamefont {Rehfeldt}}, \ and\ \bibinfo {author}
  {\bibfnamefont {Christoph}\ \bibnamefont {Westerhausen}},\ }\bibfield
  {title} {\enquote {\bibinfo {title} {The 2019 surface acoustic waves
  roadmap},}\ }\href {\doibase 10.1088/1361-6463/ab1b04} {\bibfield  {journal}
  {\bibinfo  {journal} {Journal of Physics D: Applied Physics}\ }\textbf
  {\bibinfo {volume} {52}},\ \bibinfo {pages} {353001} (\bibinfo {year}
  {2019})}\BibitemShut {NoStop}%
\bibitem [{\citenamefont {Golter}\ \emph {et~al.}(2016)\citenamefont {Golter},
  \citenamefont {Oo}, \citenamefont {Amezcua}, \citenamefont {Lekavicius},
  \citenamefont {Stewart},\ and\ \citenamefont
  {Wang}}]{golterCouplingSurfaceAcoustic2016}%
  \BibitemOpen
  \bibfield  {author} {\bibinfo {author} {\bibfnamefont {D.~Andrew}\
  \bibnamefont {Golter}}, \bibinfo {author} {\bibfnamefont {Thein}\
  \bibnamefont {Oo}}, \bibinfo {author} {\bibfnamefont {Mayra}\ \bibnamefont
  {Amezcua}}, \bibinfo {author} {\bibfnamefont {Ignas}\ \bibnamefont
  {Lekavicius}}, \bibinfo {author} {\bibfnamefont {Kevin~A.}\ \bibnamefont
  {Stewart}}, \ and\ \bibinfo {author} {\bibfnamefont {Hailin}\ \bibnamefont
  {Wang}},\ }\bibfield  {title} {\enquote {\bibinfo {title} {Coupling a
  {{Surface Acoustic Wave}} to an {{Electron Spin}} in {{Diamond}} via a {{Dark
  State}}},}\ }\href {\doibase 10.1103/PhysRevX.6.041060} {\bibfield  {journal}
  {\bibinfo  {journal} {Physical Review X}\ }\textbf {\bibinfo {volume} {6}},\
  \bibinfo {pages} {041060} (\bibinfo {year} {2016})}\BibitemShut {NoStop}%
\bibitem [{\citenamefont {Whiteley}\ \emph {et~al.}(2019)\citenamefont
  {Whiteley}, \citenamefont {Wolfowicz}, \citenamefont {Anderson},
  \citenamefont {Bourassa}, \citenamefont {Ma}, \citenamefont {Ye},
  \citenamefont {Koolstra}, \citenamefont {Satzinger}, \citenamefont {Holt},
  \citenamefont {Heremans}, \citenamefont {Cleland}, \citenamefont {Schuster},
  \citenamefont {Galli},\ and\ \citenamefont
  {Awschalom}}]{whiteleySpinPhononInteractions2019}%
  \BibitemOpen
  \bibfield  {author} {\bibinfo {author} {\bibfnamefont {Samuel~J.}\
  \bibnamefont {Whiteley}}, \bibinfo {author} {\bibfnamefont {Gary}\
  \bibnamefont {Wolfowicz}}, \bibinfo {author} {\bibfnamefont {Christopher~P.}\
  \bibnamefont {Anderson}}, \bibinfo {author} {\bibfnamefont {Alexandre}\
  \bibnamefont {Bourassa}}, \bibinfo {author} {\bibfnamefont {He}~\bibnamefont
  {Ma}}, \bibinfo {author} {\bibfnamefont {Meng}\ \bibnamefont {Ye}}, \bibinfo
  {author} {\bibfnamefont {Gerwin}\ \bibnamefont {Koolstra}}, \bibinfo {author}
  {\bibfnamefont {Kevin~J.}\ \bibnamefont {Satzinger}}, \bibinfo {author}
  {\bibfnamefont {Martin~V.}\ \bibnamefont {Holt}}, \bibinfo {author}
  {\bibfnamefont {F.~Joseph}\ \bibnamefont {Heremans}}, \bibinfo {author}
  {\bibfnamefont {Andrew~N.}\ \bibnamefont {Cleland}}, \bibinfo {author}
  {\bibfnamefont {David~I.}\ \bibnamefont {Schuster}}, \bibinfo {author}
  {\bibfnamefont {Giulia}\ \bibnamefont {Galli}}, \ and\ \bibinfo {author}
  {\bibfnamefont {David~D.}\ \bibnamefont {Awschalom}},\ }\bibfield  {title}
  {\enquote {\bibinfo {title} {Spin\textendash{}phonon interactions in silicon
  carbide addressed by {{Gaussian}} acoustics},}\ }\href {\doibase
  10.1038/s41567-019-0420-0} {\bibfield  {journal} {\bibinfo  {journal} {Nature
  Physics}\ ,\ \bibinfo {pages} {1}} (\bibinfo {year} {2019})}\BibitemShut
  {NoStop}%
\bibitem [{\citenamefont {Bochmann}\ \emph {et~al.}(2013)\citenamefont
  {Bochmann}, \citenamefont {Vainsencher}, \citenamefont {Awschalom},\ and\
  \citenamefont {Cleland}}]{bochmann2013}%
  \BibitemOpen
  \bibfield  {author} {\bibinfo {author} {\bibfnamefont {J.}~\bibnamefont
  {Bochmann}}, \bibinfo {author} {\bibfnamefont {A.}~\bibnamefont
  {Vainsencher}}, \bibinfo {author} {\bibfnamefont {D.D.}\ \bibnamefont
  {Awschalom}}, \ and\ \bibinfo {author} {\bibfnamefont {A.N.}\ \bibnamefont
  {Cleland}},\ }\bibfield  {title} {\enquote {\bibinfo {title} {Nanomechanical
  coupling between microwave and optical photons},}\ }\href {\doibase
  10.1088/1361-6463/ab1b04} {\bibfield  {journal} {\bibinfo  {journal} {Nature
  Physics}\ }\textbf {\bibinfo {volume} {9}},\ \bibinfo {pages} {712--716}
  (\bibinfo {year} {2013})}\BibitemShut {NoStop}%
\bibitem [{\citenamefont {Vainsencher}\ \emph {et~al.}(2016)\citenamefont
  {Vainsencher}, \citenamefont {Satzinger}, \citenamefont {Peairs},\ and\
  \citenamefont {Cleland}}]{vainsencherBidirectionalConversionMicrowave2016}%
  \BibitemOpen
  \bibfield  {author} {\bibinfo {author} {\bibfnamefont {Amit}\ \bibnamefont
  {Vainsencher}}, \bibinfo {author} {\bibfnamefont {K.~J.}\ \bibnamefont
  {Satzinger}}, \bibinfo {author} {\bibfnamefont {G.~A.}\ \bibnamefont
  {Peairs}}, \ and\ \bibinfo {author} {\bibfnamefont {A.~N.}\ \bibnamefont
  {Cleland}},\ }\bibfield  {title} {\enquote {\bibinfo {title} {Bi-directional
  conversion between microwave and optical frequencies in a piezoelectric
  optomechanical device},}\ }\href {\doibase 10.1063/1.4955408} {\bibfield
  {journal} {\bibinfo  {journal} {Applied Physics Letters}\ }\textbf {\bibinfo
  {volume} {109}},\ \bibinfo {pages} {033107} (\bibinfo {year}
  {2016})}\BibitemShut {NoStop}%
\bibitem [{\citenamefont
  {Shumeiko}(2016)}]{shumeikoQuantumAcoustoopticTransducer2016}%
  \BibitemOpen
  \bibfield  {author} {\bibinfo {author} {\bibfnamefont {Vitaly~S.}\
  \bibnamefont {Shumeiko}},\ }\bibfield  {title} {\enquote {\bibinfo {title}
  {Quantum acousto-optic transducer for superconducting qubits},}\ }\href
  {\doibase 10.1103/PhysRevA.93.023838} {\bibfield  {journal} {\bibinfo
  {journal} {Physical Review A}\ }\textbf {\bibinfo {volume} {93}},\ \bibinfo
  {pages} {023838} (\bibinfo {year} {2016})}\BibitemShut {NoStop}%
\bibitem [{\citenamefont {McNeil}\ \emph {et~al.}(2011)\citenamefont {McNeil},
  \citenamefont {Kataoka}, \citenamefont {Ford}, \citenamefont {Barnes},
  \citenamefont {Anderson}, \citenamefont {Jones}, \citenamefont {Farrer},\
  and\ \citenamefont {Ritchie}}]{mcneilOndemandSingleelectronTransfer2011}%
  \BibitemOpen
  \bibfield  {author} {\bibinfo {author} {\bibfnamefont {R.~P.~G.}\
  \bibnamefont {McNeil}}, \bibinfo {author} {\bibfnamefont {M.}~\bibnamefont
  {Kataoka}}, \bibinfo {author} {\bibfnamefont {C.~J.~B.}\ \bibnamefont
  {Ford}}, \bibinfo {author} {\bibfnamefont {C.~H.~W.}\ \bibnamefont {Barnes}},
  \bibinfo {author} {\bibfnamefont {D.}~\bibnamefont {Anderson}}, \bibinfo
  {author} {\bibfnamefont {G.~a.~C.}\ \bibnamefont {Jones}}, \bibinfo {author}
  {\bibfnamefont {I.}~\bibnamefont {Farrer}}, \ and\ \bibinfo {author}
  {\bibfnamefont {D.~A.}\ \bibnamefont {Ritchie}},\ }\bibfield  {title}
  {\enquote {\bibinfo {title} {On-demand single-electron transfer between
  distant quantum dots},}\ }\href {\doibase 10.1038/nature10444} {\bibfield
  {journal} {\bibinfo  {journal} {Nature}\ }\textbf {\bibinfo {volume} {477}},\
  \bibinfo {pages} {439--442} (\bibinfo {year} {2011})}\BibitemShut {NoStop}%
\bibitem [{\citenamefont {Hermelin}\ \emph {et~al.}(2011)\citenamefont
  {Hermelin}, \citenamefont {Takada}, \citenamefont {Yamamoto}, \citenamefont
  {Tarucha}, \citenamefont {Wieck}, \citenamefont {Saminadayar}, \citenamefont
  {B{\"a}uerle},\ and\ \citenamefont
  {Meunier}}]{hermelinElectronsSurfingSound2011}%
  \BibitemOpen
  \bibfield  {author} {\bibinfo {author} {\bibfnamefont {Sylvain}\ \bibnamefont
  {Hermelin}}, \bibinfo {author} {\bibfnamefont {Shintaro}\ \bibnamefont
  {Takada}}, \bibinfo {author} {\bibfnamefont {Michihisa}\ \bibnamefont
  {Yamamoto}}, \bibinfo {author} {\bibfnamefont {Seigo}\ \bibnamefont
  {Tarucha}}, \bibinfo {author} {\bibfnamefont {Andreas~D.}\ \bibnamefont
  {Wieck}}, \bibinfo {author} {\bibfnamefont {Laurent}\ \bibnamefont
  {Saminadayar}}, \bibinfo {author} {\bibfnamefont {Christopher}\ \bibnamefont
  {B{\"a}uerle}}, \ and\ \bibinfo {author} {\bibfnamefont {Tristan}\
  \bibnamefont {Meunier}},\ }\bibfield  {title} {\enquote {\bibinfo {title}
  {Electrons surfing on a sound wave as a platform for quantum optics with
  flying electrons},}\ }\href {\doibase 10.1038/nature10416} {\bibfield
  {journal} {\bibinfo  {journal} {Nature}\ }\textbf {\bibinfo {volume} {477}},\
  \bibinfo {pages} {435--438} (\bibinfo {year} {2011})}\BibitemShut {NoStop}%
\bibitem [{\citenamefont {Gustafsson}\ \emph {et~al.}(2014)\citenamefont
  {Gustafsson}, \citenamefont {Aref}, \citenamefont {Kockum}, \citenamefont
  {Ekstr{\"o}m}, \citenamefont {Johansson},\ and\ \citenamefont
  {Delsing}}]{gustafssonPropagatingPhononsCoupled2014}%
  \BibitemOpen
  \bibfield  {author} {\bibinfo {author} {\bibfnamefont {Martin~V.}\
  \bibnamefont {Gustafsson}}, \bibinfo {author} {\bibfnamefont {Thomas}\
  \bibnamefont {Aref}}, \bibinfo {author} {\bibfnamefont {Anton~Frisk}\
  \bibnamefont {Kockum}}, \bibinfo {author} {\bibfnamefont {Maria~K.}\
  \bibnamefont {Ekstr{\"o}m}}, \bibinfo {author} {\bibfnamefont {G{\"o}ran}\
  \bibnamefont {Johansson}}, \ and\ \bibinfo {author} {\bibfnamefont {Per}\
  \bibnamefont {Delsing}},\ }\bibfield  {title} {\enquote {\bibinfo {title}
  {Propagating phonons coupled to an artificial atom},}\ }\href {\doibase
  10.1126/science.1257219} {\bibfield  {journal} {\bibinfo  {journal}
  {Science}\ }\textbf {\bibinfo {volume} {346}},\ \bibinfo {pages} {207--211}
  (\bibinfo {year} {2014})}\BibitemShut {NoStop}%
\bibitem [{\citenamefont {Manenti}\ \emph {et~al.}(2017)\citenamefont
  {Manenti}, \citenamefont {Kockum}, \citenamefont {Patterson}, \citenamefont
  {Behrle}, \citenamefont {Rahamim}, \citenamefont {Tancredi}, \citenamefont
  {Nori},\ and\ \citenamefont
  {Leek}}]{manentiCircuitQuantumAcoustodynamics2017}%
  \BibitemOpen
  \bibfield  {author} {\bibinfo {author} {\bibfnamefont {Riccardo}\
  \bibnamefont {Manenti}}, \bibinfo {author} {\bibfnamefont {Anton~F.}\
  \bibnamefont {Kockum}}, \bibinfo {author} {\bibfnamefont {Andrew}\
  \bibnamefont {Patterson}}, \bibinfo {author} {\bibfnamefont {Tanja}\
  \bibnamefont {Behrle}}, \bibinfo {author} {\bibfnamefont {Joseph}\
  \bibnamefont {Rahamim}}, \bibinfo {author} {\bibfnamefont {Giovanna}\
  \bibnamefont {Tancredi}}, \bibinfo {author} {\bibfnamefont {Franco}\
  \bibnamefont {Nori}}, \ and\ \bibinfo {author} {\bibfnamefont {Peter~J.}\
  \bibnamefont {Leek}},\ }\bibfield  {title} {\enquote {\bibinfo {title}
  {Circuit quantum acoustodynamics with surface acoustic waves},}\ }\href
  {\doibase 10.1038/s41467-017-01063-9} {\bibfield  {journal} {\bibinfo
  {journal} {Nature Communications}\ }\textbf {\bibinfo {volume} {8}},\
  \bibinfo {pages} {975} (\bibinfo {year} {2017})}\BibitemShut {NoStop}%
\bibitem [{\citenamefont {Moores}\ \emph {et~al.}(2018)\citenamefont {Moores},
  \citenamefont {Sletten}, \citenamefont {Viennot},\ and\ \citenamefont
  {Lehnert}}]{mooresCavityQuantumAcoustic2018}%
  \BibitemOpen
  \bibfield  {author} {\bibinfo {author} {\bibfnamefont {Bradley~A.}\
  \bibnamefont {Moores}}, \bibinfo {author} {\bibfnamefont {Lucas~R.}\
  \bibnamefont {Sletten}}, \bibinfo {author} {\bibfnamefont {Jeremie~J.}\
  \bibnamefont {Viennot}}, \ and\ \bibinfo {author} {\bibfnamefont {K.~W.}\
  \bibnamefont {Lehnert}},\ }\bibfield  {title} {{\selectlanguage
  {English}\enquote {\bibinfo {title} {Cavity {{Quantum Acoustic Device}} in
  the {{Multimode Strong Coupling Regime}}},}\ }}\href {\doibase
  10.1103/PhysRevLett.120.227701} {\bibfield  {journal} {\bibinfo  {journal}
  {Physical Review Letters}\ }\textbf {\bibinfo {volume} {120}} (\bibinfo
  {year} {2018}),\ 10.1103/PhysRevLett.120.227701}\BibitemShut {NoStop}%
\bibitem [{\citenamefont {Bolgar}\ \emph {et~al.}(2018)\citenamefont {Bolgar},
  \citenamefont {Zotova}, \citenamefont {Kirichenko}, \citenamefont {Besedin},
  \citenamefont {Semenov}, \citenamefont {Shaikhaidarov},\ and\ \citenamefont
  {Astafiev}}]{bolgarQuantumRegimeTwoDimensional2018}%
  \BibitemOpen
  \bibfield  {author} {\bibinfo {author} {\bibfnamefont {Aleksey~N.}\
  \bibnamefont {Bolgar}}, \bibinfo {author} {\bibfnamefont {Julia~I.}\
  \bibnamefont {Zotova}}, \bibinfo {author} {\bibfnamefont {Daniil~D.}\
  \bibnamefont {Kirichenko}}, \bibinfo {author} {\bibfnamefont {Ilia~S.}\
  \bibnamefont {Besedin}}, \bibinfo {author} {\bibfnamefont {Aleksander~V.}\
  \bibnamefont {Semenov}}, \bibinfo {author} {\bibfnamefont {Rais~S.}\
  \bibnamefont {Shaikhaidarov}}, \ and\ \bibinfo {author} {\bibfnamefont
  {Oleg~V.}\ \bibnamefont {Astafiev}},\ }\bibfield  {title} {\enquote {\bibinfo
  {title} {Quantum {{Regime}} of a {{Two}}-{{Dimensional Phonon Cavity}}},}\
  }\href {\doibase 10.1103/PhysRevLett.120.223603} {\bibfield  {journal}
  {\bibinfo  {journal} {Physical Review Letters}\ }\textbf {\bibinfo {volume}
  {120}},\ \bibinfo {pages} {223603} (\bibinfo {year} {2018})}\BibitemShut
  {NoStop}%
\bibitem [{\citenamefont {Noguchi}\ \emph {et~al.}(2017)\citenamefont
  {Noguchi}, \citenamefont {Yamazaki}, \citenamefont {Tabuchi},\ and\
  \citenamefont {Nakamura}}]{noguchiQubitAssistedTransductionDetection2017}%
  \BibitemOpen
  \bibfield  {author} {\bibinfo {author} {\bibfnamefont {Atsushi}\ \bibnamefont
  {Noguchi}}, \bibinfo {author} {\bibfnamefont {Rekishu}\ \bibnamefont
  {Yamazaki}}, \bibinfo {author} {\bibfnamefont {Yutaka}\ \bibnamefont
  {Tabuchi}}, \ and\ \bibinfo {author} {\bibfnamefont {Yasunobu}\ \bibnamefont
  {Nakamura}},\ }\bibfield  {title} {\enquote {\bibinfo {title}
  {Qubit-{{Assisted Transduction}} for a {{Detection}} of {{Surface Acoustic
  Waves}} near the {{Quantum Limit}}},}\ }\href {\doibase
  10.1103/PhysRevLett.119.180505} {\bibfield  {journal} {\bibinfo  {journal}
  {Physical Review Letters}\ }\textbf {\bibinfo {volume} {119}},\ \bibinfo
  {pages} {180505} (\bibinfo {year} {2017})}\BibitemShut {NoStop}%
\bibitem [{\citenamefont {Satzinger}\ \emph {et~al.}(2018)\citenamefont
  {Satzinger}, \citenamefont {Zhong}, \citenamefont {Chang}, \citenamefont
  {Peairs}, \citenamefont {Bienfait}, \citenamefont {Chou}, \citenamefont
  {Cleland}, \citenamefont {Conner}, \citenamefont {Dumur}, \citenamefont
  {Grebel}, \citenamefont {Gutierrez}, \citenamefont {November}, \citenamefont
  {Povey}, \citenamefont {Whiteley}, \citenamefont {Awschalom}, \citenamefont
  {Schuster},\ and\ \citenamefont
  {Cleland}}]{satzingerQuantumControlSurface2018}%
  \BibitemOpen
  \bibfield  {author} {\bibinfo {author} {\bibfnamefont {K.~J.}\ \bibnamefont
  {Satzinger}}, \bibinfo {author} {\bibfnamefont {Y.~P.}\ \bibnamefont
  {Zhong}}, \bibinfo {author} {\bibfnamefont {H.-S.}\ \bibnamefont {Chang}},
  \bibinfo {author} {\bibfnamefont {G.~A.}\ \bibnamefont {Peairs}}, \bibinfo
  {author} {\bibfnamefont {A.}~\bibnamefont {Bienfait}}, \bibinfo {author}
  {\bibfnamefont {Ming-Han}\ \bibnamefont {Chou}}, \bibinfo {author}
  {\bibfnamefont {A.~Y.}\ \bibnamefont {Cleland}}, \bibinfo {author}
  {\bibfnamefont {C.~R.}\ \bibnamefont {Conner}}, \bibinfo {author}
  {\bibfnamefont {{\'E}.}~\bibnamefont {Dumur}}, \bibinfo {author}
  {\bibfnamefont {J.}~\bibnamefont {Grebel}}, \bibinfo {author} {\bibfnamefont
  {I.}~\bibnamefont {Gutierrez}}, \bibinfo {author} {\bibfnamefont {B.~H.}\
  \bibnamefont {November}}, \bibinfo {author} {\bibfnamefont {R.~G.}\
  \bibnamefont {Povey}}, \bibinfo {author} {\bibfnamefont {S.~J.}\ \bibnamefont
  {Whiteley}}, \bibinfo {author} {\bibfnamefont {D.~D.}\ \bibnamefont
  {Awschalom}}, \bibinfo {author} {\bibfnamefont {D.~I.}\ \bibnamefont
  {Schuster}}, \ and\ \bibinfo {author} {\bibfnamefont {A.~N.}\ \bibnamefont
  {Cleland}},\ }\bibfield  {title} {\enquote {\bibinfo {title} {Quantum
  {{Control}} of {{Surface Acoustic}}-{{Wave Phonons}}},}\ }\href {\doibase
  10.1038/s41586-018-0719-5} {\bibfield  {journal} {\bibinfo  {journal}
  {Nature}\ }\textbf {\bibinfo {volume} {563}},\ \bibinfo {pages} {661--665}
  (\bibinfo {year} {2018})}\BibitemShut {NoStop}%
\bibitem [{\citenamefont {Sletten}\ \emph {et~al.}(2019)\citenamefont
  {Sletten}, \citenamefont {Moores}, \citenamefont {Viennot},\ and\
  \citenamefont {Lehnert}}]{slettenResolvingPhononFock2019}%
  \BibitemOpen
  \bibfield  {author} {\bibinfo {author} {\bibfnamefont {L.~R.}\ \bibnamefont
  {Sletten}}, \bibinfo {author} {\bibfnamefont {B.~A.}\ \bibnamefont {Moores}},
  \bibinfo {author} {\bibfnamefont {J.~J.}\ \bibnamefont {Viennot}}, \ and\
  \bibinfo {author} {\bibfnamefont {K.~W.}\ \bibnamefont {Lehnert}},\
  }\bibfield  {title} {\enquote {\bibinfo {title} {Resolving {{Phonon Fock
  States}} in a {{Multimode Cavity}} with a {{Double}}-{{Slit Qubit}}},}\
  }\href {\doibase 10.1103/PhysRevX.9.021056} {\bibfield  {journal} {\bibinfo
  {journal} {Physical Review X}\ }\textbf {\bibinfo {volume} {9}},\ \bibinfo
  {pages} {021056} (\bibinfo {year} {2019})}\BibitemShut {NoStop}%
\bibitem [{\citenamefont {Andersson}\ \emph
  {et~al.}(2019{\natexlab{a}})\citenamefont {Andersson}, \citenamefont {Suri},
  \citenamefont {Guo}, \citenamefont {Aref},\ and\ \citenamefont
  {Delsing}}]{anderssonNonexponentialDecayGiant2019}%
  \BibitemOpen
  \bibfield  {author} {\bibinfo {author} {\bibfnamefont {Gustav}\ \bibnamefont
  {Andersson}}, \bibinfo {author} {\bibfnamefont {Baladitya}\ \bibnamefont
  {Suri}}, \bibinfo {author} {\bibfnamefont {Lingzhen}\ \bibnamefont {Guo}},
  \bibinfo {author} {\bibfnamefont {Thomas}\ \bibnamefont {Aref}}, \ and\
  \bibinfo {author} {\bibfnamefont {Per}\ \bibnamefont {Delsing}},\ }\bibfield
  {title} {\enquote {\bibinfo {title} {Non-exponential decay of a giant
  artificial atom},}\ }\href {\doibase 10.1038/s41567-019-0605-6} {\bibfield
  {journal} {\bibinfo  {journal} {Nature Physics}\ ,\ \bibinfo {pages} {1--5}}
  (\bibinfo {year} {2019}{\natexlab{a}})}\BibitemShut {NoStop}%
\bibitem [{\citenamefont {Ekstr{\"o}m}\ \emph {et~al.}(2019)\citenamefont
  {Ekstr{\"o}m}, \citenamefont {Aref}, \citenamefont {Ask}, \citenamefont
  {Andersson}, \citenamefont {Suri}, \citenamefont {Sanada}, \citenamefont
  {Johansson},\ and\ \citenamefont
  {Delsing}}]{ekstromPhononRoutingControlling2019}%
  \BibitemOpen
  \bibfield  {author} {\bibinfo {author} {\bibfnamefont {M.~K.}\ \bibnamefont
  {Ekstr{\"o}m}}, \bibinfo {author} {\bibfnamefont {T.}~\bibnamefont {Aref}},
  \bibinfo {author} {\bibfnamefont {A.}~\bibnamefont {Ask}}, \bibinfo {author}
  {\bibfnamefont {G.}~\bibnamefont {Andersson}}, \bibinfo {author}
  {\bibfnamefont {B.}~\bibnamefont {Suri}}, \bibinfo {author} {\bibfnamefont
  {H.}~\bibnamefont {Sanada}}, \bibinfo {author} {\bibfnamefont
  {G.}~\bibnamefont {Johansson}}, \ and\ \bibinfo {author} {\bibfnamefont
  {P.}~\bibnamefont {Delsing}},\ }\bibfield  {title} {\enquote {\bibinfo
  {title} {Towards phonon routing: {{Controlling}} propagating acoustic waves
  in the quantum regime},}\ }\href@noop {} {\bibfield  {journal} {\bibinfo
  {journal} {arXiv:1909.07027 [cond-mat, physics:quant-ph]}\ } (\bibinfo {year}
  {2019})},\ \Eprint {http://arxiv.org/abs/1909.07027} {arXiv:1909.07027
  [cond-mat, physics:quant-ph]} \BibitemShut {NoStop}%
\bibitem [{\citenamefont {Andersson}\ \emph
  {et~al.}(2019{\natexlab{b}})\citenamefont {Andersson}, \citenamefont
  {Ekstr{\"o}m},\ and\ \citenamefont
  {Delsing}}]{anderssonElectromagneticallyInducedTransparency2019}%
  \BibitemOpen
  \bibfield  {author} {\bibinfo {author} {\bibfnamefont {Gustav}\ \bibnamefont
  {Andersson}}, \bibinfo {author} {\bibfnamefont {Maria~K.}\ \bibnamefont
  {Ekstr{\"o}m}}, \ and\ \bibinfo {author} {\bibfnamefont {Per}\ \bibnamefont
  {Delsing}},\ }\bibfield  {title} {\enquote {\bibinfo {title}
  {Electromagnetically induced transparency in a propagating mechanical
  mode},}\ }\href@noop {} {\bibfield  {journal} {\bibinfo  {journal}
  {arXiv:1912.00777 [quant-ph]}\ } (\bibinfo {year} {2019}{\natexlab{b}})},\
  \Eprint {http://arxiv.org/abs/1912.00777} {arXiv:1912.00777 [quant-ph]}
  \BibitemShut {NoStop}%
\bibitem [{\citenamefont {Koch}\ \emph {et~al.}(2007)\citenamefont {Koch},
  \citenamefont {Yu}, \citenamefont {Gambetta}, \citenamefont {Houck},
  \citenamefont {Schuster}, \citenamefont {Majer}, \citenamefont {Blais},
  \citenamefont {Devoret}, \citenamefont {Girvin},\ and\ \citenamefont
  {Schoelkopf}}]{kochChargeinsensitiveQubitDesign2007}%
  \BibitemOpen
  \bibfield  {author} {\bibinfo {author} {\bibfnamefont {J.}~\bibnamefont
  {Koch}}, \bibinfo {author} {\bibfnamefont {T.~M.}\ \bibnamefont {Yu}},
  \bibinfo {author} {\bibfnamefont {J.}~\bibnamefont {Gambetta}}, \bibinfo
  {author} {\bibfnamefont {A.~A.}\ \bibnamefont {Houck}}, \bibinfo {author}
  {\bibfnamefont {D.~I.}\ \bibnamefont {Schuster}}, \bibinfo {author}
  {\bibfnamefont {J.}~\bibnamefont {Majer}}, \bibinfo {author} {\bibfnamefont
  {A.}~\bibnamefont {Blais}}, \bibinfo {author} {\bibfnamefont {M.~H.}\
  \bibnamefont {Devoret}}, \bibinfo {author} {\bibfnamefont {S.~M.}\
  \bibnamefont {Girvin}}, \ and\ \bibinfo {author} {\bibfnamefont {R.~J.}\
  \bibnamefont {Schoelkopf}},\ }\bibfield  {title} {\enquote {\bibinfo {title}
  {Charge-insensitive qubit design derived from the {{Cooper}} pair box},}\
  }\href {\doibase 10.1103/PhysRevA.76.042319} {\bibfield  {journal} {\bibinfo
  {journal} {Phys. Rev. A}\ }\textbf {\bibinfo {volume} {76}},\ \bibinfo
  {pages} {042319} (\bibinfo {year} {2007})}\BibitemShut {NoStop}%
\bibitem [{\citenamefont {Barends}\ \emph {et~al.}(2013)\citenamefont
  {Barends}, \citenamefont {Kelly}, \citenamefont {Megrant}, \citenamefont
  {Sank}, \citenamefont {Jeffrey}, \citenamefont {Chen}, \citenamefont {Yin},
  \citenamefont {Chiaro}, \citenamefont {Mutus}, \citenamefont {Neill},
  \citenamefont {O'Malley}, \citenamefont {Roushan}, \citenamefont {Wenner},
  \citenamefont {White}, \citenamefont {Cleland},\ and\ \citenamefont
  {Martinis}}]{barendsCoherentJosephsonQubit2013}%
  \BibitemOpen
  \bibfield  {author} {\bibinfo {author} {\bibfnamefont {R.}~\bibnamefont
  {Barends}}, \bibinfo {author} {\bibfnamefont {J.}~\bibnamefont {Kelly}},
  \bibinfo {author} {\bibfnamefont {A.}~\bibnamefont {Megrant}}, \bibinfo
  {author} {\bibfnamefont {D.}~\bibnamefont {Sank}}, \bibinfo {author}
  {\bibfnamefont {E.}~\bibnamefont {Jeffrey}}, \bibinfo {author} {\bibfnamefont
  {Y.}~\bibnamefont {Chen}}, \bibinfo {author} {\bibfnamefont {Y.}~\bibnamefont
  {Yin}}, \bibinfo {author} {\bibfnamefont {B.}~\bibnamefont {Chiaro}},
  \bibinfo {author} {\bibfnamefont {J.}~\bibnamefont {Mutus}}, \bibinfo
  {author} {\bibfnamefont {C.}~\bibnamefont {Neill}}, \bibinfo {author}
  {\bibfnamefont {P.}~\bibnamefont {O'Malley}}, \bibinfo {author}
  {\bibfnamefont {P.}~\bibnamefont {Roushan}}, \bibinfo {author} {\bibfnamefont
  {J.}~\bibnamefont {Wenner}}, \bibinfo {author} {\bibfnamefont {T.~C.}\
  \bibnamefont {White}}, \bibinfo {author} {\bibfnamefont {A.~N.}\ \bibnamefont
  {Cleland}}, \ and\ \bibinfo {author} {\bibfnamefont {John~M.}\ \bibnamefont
  {Martinis}},\ }\bibfield  {title} {\enquote {\bibinfo {title} {Coherent
  {{Josephson Qubit Suitable}} for {{Scalable Quantum Integrated Circuits}}},}\
  }\href {\doibase 10.1103/PhysRevLett.111.080502} {\bibfield  {journal}
  {\bibinfo  {journal} {Phys. Rev. Lett.}\ }\textbf {\bibinfo {volume} {111}},\
  \bibinfo {pages} {080502} (\bibinfo {year} {2013})}\BibitemShut {NoStop}%
\bibitem [{\citenamefont {Chen}\ \emph {et~al.}(2014)\citenamefont {Chen},
  \citenamefont {Neill}, \citenamefont {Roushan}, \citenamefont {Leung},
  \citenamefont {Fang}, \citenamefont {Barends}, \citenamefont {Kelly},
  \citenamefont {Campbell}, \citenamefont {Chen}, \citenamefont {Chiaro},
  \citenamefont {Dunsworth}, \citenamefont {Jeffrey}, \citenamefont {Megrant},
  \citenamefont {Mutus}, \citenamefont {O'Malley}, \citenamefont {Quintana},
  \citenamefont {Sank}, \citenamefont {Vainsencher}, \citenamefont {Wenner},
  \citenamefont {White}, \citenamefont {Geller}, \citenamefont {Cleland},\ and\
  \citenamefont {Martinis}}]{chenQubitArchitectureHigh2014}%
  \BibitemOpen
  \bibfield  {author} {\bibinfo {author} {\bibfnamefont {Yu}~\bibnamefont
  {Chen}}, \bibinfo {author} {\bibfnamefont {C.}~\bibnamefont {Neill}},
  \bibinfo {author} {\bibfnamefont {P.}~\bibnamefont {Roushan}}, \bibinfo
  {author} {\bibfnamefont {N.}~\bibnamefont {Leung}}, \bibinfo {author}
  {\bibfnamefont {M.}~\bibnamefont {Fang}}, \bibinfo {author} {\bibfnamefont
  {R.}~\bibnamefont {Barends}}, \bibinfo {author} {\bibfnamefont
  {J.}~\bibnamefont {Kelly}}, \bibinfo {author} {\bibfnamefont
  {B.}~\bibnamefont {Campbell}}, \bibinfo {author} {\bibfnamefont
  {Z.}~\bibnamefont {Chen}}, \bibinfo {author} {\bibfnamefont {B.}~\bibnamefont
  {Chiaro}}, \bibinfo {author} {\bibfnamefont {A.}~\bibnamefont {Dunsworth}},
  \bibinfo {author} {\bibfnamefont {E.}~\bibnamefont {Jeffrey}}, \bibinfo
  {author} {\bibfnamefont {A.}~\bibnamefont {Megrant}}, \bibinfo {author}
  {\bibfnamefont {J.~Y.}\ \bibnamefont {Mutus}}, \bibinfo {author}
  {\bibfnamefont {P.~J.~J.}\ \bibnamefont {O'Malley}}, \bibinfo {author}
  {\bibfnamefont {C.~M.}\ \bibnamefont {Quintana}}, \bibinfo {author}
  {\bibfnamefont {D.}~\bibnamefont {Sank}}, \bibinfo {author} {\bibfnamefont
  {A.}~\bibnamefont {Vainsencher}}, \bibinfo {author} {\bibfnamefont
  {J.}~\bibnamefont {Wenner}}, \bibinfo {author} {\bibfnamefont {T.~C.}\
  \bibnamefont {White}}, \bibinfo {author} {\bibfnamefont {Michael~R.}\
  \bibnamefont {Geller}}, \bibinfo {author} {\bibfnamefont {A.~N.}\
  \bibnamefont {Cleland}}, \ and\ \bibinfo {author} {\bibfnamefont {John~M.}\
  \bibnamefont {Martinis}},\ }\bibfield  {title} {\enquote {\bibinfo {title}
  {Qubit {{Architecture}} with {{High Coherence}} and {{Fast Tunable
  Coupling}}},}\ }\href {\doibase 10.1103/PhysRevLett.113.220502} {\bibfield
  {journal} {\bibinfo  {journal} {Physical Review Letters}\ }\textbf {\bibinfo
  {volume} {113}},\ \bibinfo {pages} {220502} (\bibinfo {year}
  {2014})}\BibitemShut {NoStop}%
\bibitem [{\citenamefont {Zhong}\ \emph {et~al.}(2019)\citenamefont {Zhong},
  \citenamefont {Chang}, \citenamefont {Satzinger}, \citenamefont {Chou},
  \citenamefont {Bienfait}, \citenamefont {Conner}, \citenamefont {Dumur},
  \citenamefont {Grebel}, \citenamefont {Peairs}, \citenamefont {Povey},
  \citenamefont {Schuster},\ and\ \citenamefont
  {Cleland}}]{zhongViolatingBellInequality2018}%
  \BibitemOpen
  \bibfield  {author} {\bibinfo {author} {\bibfnamefont {Y.~P.}\ \bibnamefont
  {Zhong}}, \bibinfo {author} {\bibfnamefont {H.-S.}\ \bibnamefont {Chang}},
  \bibinfo {author} {\bibfnamefont {K.~J.}\ \bibnamefont {Satzinger}}, \bibinfo
  {author} {\bibfnamefont {M.-H.}\ \bibnamefont {Chou}}, \bibinfo {author}
  {\bibfnamefont {A.}~\bibnamefont {Bienfait}}, \bibinfo {author}
  {\bibfnamefont {C.~R.}\ \bibnamefont {Conner}}, \bibinfo {author}
  {\bibfnamefont {{\'E}.}~\bibnamefont {Dumur}}, \bibinfo {author}
  {\bibfnamefont {J.}~\bibnamefont {Grebel}}, \bibinfo {author} {\bibfnamefont
  {G.~A.}\ \bibnamefont {Peairs}}, \bibinfo {author} {\bibfnamefont {R.~G.}\
  \bibnamefont {Povey}}, \bibinfo {author} {\bibfnamefont {D.~I.}\ \bibnamefont
  {Schuster}}, \ and\ \bibinfo {author} {\bibfnamefont {A.~N.}\ \bibnamefont
  {Cleland}},\ }\bibfield  {title} {\enquote {\bibinfo {title} {Violating
  {{Bell}}'s inequality with remotely connected superconducting qubits},}\
  }\href {\doibase 10.1038/s41567-019-0507-7} {\bibfield  {journal} {\bibinfo
  {journal} {Nature Physics}\ }\textbf {\bibinfo {volume} {15}},\ \bibinfo
  {pages} {741--744} (\bibinfo {year} {2019})}\BibitemShut {NoStop}%
\bibitem [{\citenamefont {Korotkov}(2011)}]{korotkovFlyingMicrowaveQubits2011}%
  \BibitemOpen
  \bibfield  {author} {\bibinfo {author} {\bibfnamefont {Alexander~N.}\
  \bibnamefont {Korotkov}},\ }\bibfield  {title} {\enquote {\bibinfo {title}
  {Flying microwave qubits with nearly perfect transfer efficiency},}\ }\href
  {\doibase 10.1103/PhysRevB.84.014510} {\bibfield  {journal} {\bibinfo
  {journal} {Physical Review B}\ }\textbf {\bibinfo {volume} {84}},\ \bibinfo
  {pages} {014510} (\bibinfo {year} {2011})}\BibitemShut {NoStop}%
\bibitem [{\citenamefont {Gambetta}\ \emph {et~al.}(2011)\citenamefont
  {Gambetta}, \citenamefont {Houck},\ and\ \citenamefont
  {Blais}}]{gambettaSuperconductingQubitPurcell2011}%
  \BibitemOpen
  \bibfield  {author} {\bibinfo {author} {\bibfnamefont {J.~M.}\ \bibnamefont
  {Gambetta}}, \bibinfo {author} {\bibfnamefont {A.~A.}\ \bibnamefont {Houck}},
  \ and\ \bibinfo {author} {\bibfnamefont {Alexandre}\ \bibnamefont {Blais}},\
  }\bibfield  {title} {\enquote {\bibinfo {title} {Superconducting {{Qubit}}
  with {{Purcell Protection}} and {{Tunable Coupling}}},}\ }\href {\doibase
  10.1103/PhysRevLett.106.030502} {\bibfield  {journal} {\bibinfo  {journal}
  {Physical Review Letters}\ }\textbf {\bibinfo {volume} {106}},\ \bibinfo
  {pages} {030502} (\bibinfo {year} {2011})}\BibitemShut {NoStop}%
\bibitem [{\citenamefont {Jeffrey}\ \emph {et~al.}(2014)\citenamefont
  {Jeffrey}, \citenamefont {Sank}, \citenamefont {Mutus}, \citenamefont
  {White}, \citenamefont {Kelly}, \citenamefont {Barends}, \citenamefont
  {Chen}, \citenamefont {Chen}, \citenamefont {Chiaro}, \citenamefont
  {Dunsworth}, \citenamefont {Megrant}, \citenamefont {O'Malley}, \citenamefont
  {Neill}, \citenamefont {Roushan}, \citenamefont {Vainsencher}, \citenamefont
  {Wenner}, \citenamefont {Cleland},\ and\ \citenamefont
  {Martinis}}]{jeffreyFastAccurateState2014}%
  \BibitemOpen
  \bibfield  {author} {\bibinfo {author} {\bibfnamefont {Evan}\ \bibnamefont
  {Jeffrey}}, \bibinfo {author} {\bibfnamefont {Daniel}\ \bibnamefont {Sank}},
  \bibinfo {author} {\bibfnamefont {J.~Y.}\ \bibnamefont {Mutus}}, \bibinfo
  {author} {\bibfnamefont {T.~C.}\ \bibnamefont {White}}, \bibinfo {author}
  {\bibfnamefont {J.}~\bibnamefont {Kelly}}, \bibinfo {author} {\bibfnamefont
  {R.}~\bibnamefont {Barends}}, \bibinfo {author} {\bibfnamefont
  {Y.}~\bibnamefont {Chen}}, \bibinfo {author} {\bibfnamefont {Z.}~\bibnamefont
  {Chen}}, \bibinfo {author} {\bibfnamefont {B.}~\bibnamefont {Chiaro}},
  \bibinfo {author} {\bibfnamefont {A.}~\bibnamefont {Dunsworth}}, \bibinfo
  {author} {\bibfnamefont {A.}~\bibnamefont {Megrant}}, \bibinfo {author}
  {\bibfnamefont {P.~J.~J.}\ \bibnamefont {O'Malley}}, \bibinfo {author}
  {\bibfnamefont {C.}~\bibnamefont {Neill}}, \bibinfo {author} {\bibfnamefont
  {P.}~\bibnamefont {Roushan}}, \bibinfo {author} {\bibfnamefont
  {A.}~\bibnamefont {Vainsencher}}, \bibinfo {author} {\bibfnamefont
  {J.}~\bibnamefont {Wenner}}, \bibinfo {author} {\bibfnamefont {A.~N.}\
  \bibnamefont {Cleland}}, \ and\ \bibinfo {author} {\bibfnamefont {John~M.}\
  \bibnamefont {Martinis}},\ }\bibfield  {title} {\enquote {\bibinfo {title}
  {Fast {{Accurate State Measurement}} with {{Superconducting Qubits}}},}\
  }\href {\doibase 10.1103/PhysRevLett.112.190504} {\bibfield  {journal}
  {\bibinfo  {journal} {Physical Review Letters}\ }\textbf {\bibinfo {volume}
  {112}},\ \bibinfo {pages} {190504} (\bibinfo {year} {2014})}\BibitemShut
  {NoStop}%
\bibitem [{\citenamefont {Hoi}\ \emph {et~al.}(2015)\citenamefont {Hoi},
  \citenamefont {Kockum}, \citenamefont {Tornberg}, \citenamefont
  {Pourkabirian}, \citenamefont {Johansson}, \citenamefont {Delsing},\ and\
  \citenamefont {Wilson}}]{hoiProbingQuantumVacuum2015}%
  \BibitemOpen
  \bibfield  {author} {\bibinfo {author} {\bibfnamefont {I.-C.}\ \bibnamefont
  {Hoi}}, \bibinfo {author} {\bibfnamefont {A.~F.}\ \bibnamefont {Kockum}},
  \bibinfo {author} {\bibfnamefont {L.}~\bibnamefont {Tornberg}}, \bibinfo
  {author} {\bibfnamefont {A.}~\bibnamefont {Pourkabirian}}, \bibinfo {author}
  {\bibfnamefont {G.}~\bibnamefont {Johansson}}, \bibinfo {author}
  {\bibfnamefont {P.}~\bibnamefont {Delsing}}, \ and\ \bibinfo {author}
  {\bibfnamefont {C.~M.}\ \bibnamefont {Wilson}},\ }\bibfield  {title}
  {\enquote {\bibinfo {title} {Probing the quantum vacuum with an artificial
  atom in front of a mirror},}\ }\href {\doibase 10.1038/nphys3484} {\bibfield
  {journal} {\bibinfo  {journal} {Nature Physics}\ }\textbf {\bibinfo {volume}
  {11}},\ \bibinfo {pages} {1045--1049} (\bibinfo {year} {2015})}\BibitemShut
  {NoStop}%
\bibitem [{\citenamefont {Pfaff}\ \emph {et~al.}(2017)\citenamefont {Pfaff},
  \citenamefont {Axline}, \citenamefont {Burkhart}, \citenamefont {Vool},
  \citenamefont {Reinhold}, \citenamefont {Frunzio}, \citenamefont {Jiang},
  \citenamefont {Devoret},\ and\ \citenamefont
  {Schoelkopf}}]{pfaffControlledReleaseMultiphoton2017}%
  \BibitemOpen
  \bibfield  {author} {\bibinfo {author} {\bibfnamefont {Wolfgang}\
  \bibnamefont {Pfaff}}, \bibinfo {author} {\bibfnamefont {Christopher~J.}\
  \bibnamefont {Axline}}, \bibinfo {author} {\bibfnamefont {Luke~D.}\
  \bibnamefont {Burkhart}}, \bibinfo {author} {\bibfnamefont {Uri}\
  \bibnamefont {Vool}}, \bibinfo {author} {\bibfnamefont {Philip}\ \bibnamefont
  {Reinhold}}, \bibinfo {author} {\bibfnamefont {Luigi}\ \bibnamefont
  {Frunzio}}, \bibinfo {author} {\bibfnamefont {Liang}\ \bibnamefont {Jiang}},
  \bibinfo {author} {\bibfnamefont {Michel~H.}\ \bibnamefont {Devoret}}, \ and\
  \bibinfo {author} {\bibfnamefont {Robert~J.}\ \bibnamefont {Schoelkopf}},\
  }\bibfield  {title} {\enquote {\bibinfo {title} {Controlled release of
  multiphoton quantum states from a microwave cavity memory},}\ }\href
  {\doibase 10.1038/nphys4143} {\bibfield  {journal} {\bibinfo  {journal}
  {Nature Physics}\ }\textbf {\bibinfo {volume} {13}},\ \bibinfo {pages}
  {882--887} (\bibinfo {year} {2017})}\BibitemShut {NoStop}%
\bibitem [{\citenamefont {Frisk~Kockum}\ \emph {et~al.}(2014)\citenamefont
  {Frisk~Kockum}, \citenamefont {Delsing},\ and\ \citenamefont
  {Johansson}}]{friskkockumDesigningFrequencydependentRelaxation2014}%
  \BibitemOpen
  \bibfield  {author} {\bibinfo {author} {\bibfnamefont {Anton}\ \bibnamefont
  {Frisk~Kockum}}, \bibinfo {author} {\bibfnamefont {Per}\ \bibnamefont
  {Delsing}}, \ and\ \bibinfo {author} {\bibfnamefont {G{\"o}ran}\ \bibnamefont
  {Johansson}},\ }\bibfield  {title} {\enquote {\bibinfo {title} {Designing
  frequency-dependent relaxation rates and {{Lamb}} shifts for a giant
  artificial atom},}\ }\href {\doibase 10.1103/PhysRevA.90.013837} {\bibfield
  {journal} {\bibinfo  {journal} {Physical Review A}\ }\textbf {\bibinfo
  {volume} {90}},\ \bibinfo {pages} {013837} (\bibinfo {year}
  {2014})}\BibitemShut {NoStop}%
\bibitem [{\citenamefont {Material}()}]{SupplementaryMaterial}%
  \BibitemOpen
  \bibfield  {author} {\bibinfo {author} {\bibfnamefont {Supplementary}\
  \bibnamefont {Material}},\ }\bibfield  {title} {\enquote {\bibinfo {title}
  {{{SupplementaryMaterial}}},}\ }\href@noop {} {\ }\BibitemShut {NoStop}%
\bibitem [{\citenamefont {Nigg}\ \emph {et~al.}(2012)\citenamefont {Nigg},
  \citenamefont {Paik}, \citenamefont {Vlastakis}, \citenamefont {Kirchmair},
  \citenamefont {Shankar}, \citenamefont {Frunzio}, \citenamefont {Devoret},
  \citenamefont {Schoelkopf},\ and\ \citenamefont
  {Girvin}}]{niggBlackBoxSuperconductingCircuit2012}%
  \BibitemOpen
  \bibfield  {author} {\bibinfo {author} {\bibfnamefont {Simon~E.}\
  \bibnamefont {Nigg}}, \bibinfo {author} {\bibfnamefont {Hanhee}\ \bibnamefont
  {Paik}}, \bibinfo {author} {\bibfnamefont {Brian}\ \bibnamefont {Vlastakis}},
  \bibinfo {author} {\bibfnamefont {Gerhard}\ \bibnamefont {Kirchmair}},
  \bibinfo {author} {\bibfnamefont {S.}~\bibnamefont {Shankar}}, \bibinfo
  {author} {\bibfnamefont {Luigi}\ \bibnamefont {Frunzio}}, \bibinfo {author}
  {\bibfnamefont {M.~H.}\ \bibnamefont {Devoret}}, \bibinfo {author}
  {\bibfnamefont {R.~J.}\ \bibnamefont {Schoelkopf}}, \ and\ \bibinfo {author}
  {\bibfnamefont {S.~M.}\ \bibnamefont {Girvin}},\ }\bibfield  {title}
  {\enquote {\bibinfo {title} {Black-{{Box Superconducting Circuit
  Quantization}}},}\ }\href {\doibase 10.1103/PhysRevLett.108.240502}
  {\bibfield  {journal} {\bibinfo  {journal} {Physical Review Letters}\
  }\textbf {\bibinfo {volume} {108}},\ \bibinfo {pages} {240502} (\bibinfo
  {year} {2012})}\BibitemShut {NoStop}%
\bibitem [{\citenamefont {{Campagne-Ibarcq}}\ \emph {et~al.}(2018)\citenamefont
  {{Campagne-Ibarcq}}, \citenamefont {{Zalys-Geller}}, \citenamefont {Narla},
  \citenamefont {Shankar}, \citenamefont {Reinhold}, \citenamefont {Burkhart},
  \citenamefont {Axline}, \citenamefont {Pfaff}, \citenamefont {Frunzio},
  \citenamefont {Schoelkopf},\ and\ \citenamefont
  {Devoret}}]{campagne-ibarcqDeterministicRemoteEntanglement2018}%
  \BibitemOpen
  \bibfield  {author} {\bibinfo {author} {\bibfnamefont {P.}~\bibnamefont
  {{Campagne-Ibarcq}}}, \bibinfo {author} {\bibfnamefont {E.}~\bibnamefont
  {{Zalys-Geller}}}, \bibinfo {author} {\bibfnamefont {A.}~\bibnamefont
  {Narla}}, \bibinfo {author} {\bibfnamefont {S.}~\bibnamefont {Shankar}},
  \bibinfo {author} {\bibfnamefont {P.}~\bibnamefont {Reinhold}}, \bibinfo
  {author} {\bibfnamefont {L.}~\bibnamefont {Burkhart}}, \bibinfo {author}
  {\bibfnamefont {C.}~\bibnamefont {Axline}}, \bibinfo {author} {\bibfnamefont
  {W.}~\bibnamefont {Pfaff}}, \bibinfo {author} {\bibfnamefont
  {L.}~\bibnamefont {Frunzio}}, \bibinfo {author} {\bibfnamefont {R.~J.}\
  \bibnamefont {Schoelkopf}}, \ and\ \bibinfo {author} {\bibfnamefont {M.~H.}\
  \bibnamefont {Devoret}},\ }\bibfield  {title} {\enquote {\bibinfo {title}
  {Deterministic {{Remote Entanglement}} of {{Superconducting Circuits}}
  through {{Microwave Two}}-{{Photon Transitions}}},}\ }\href {\doibase
  10.1103/PhysRevLett.120.200501} {\bibfield  {journal} {\bibinfo  {journal}
  {Physical Review Letters}\ }\textbf {\bibinfo {volume} {120}},\ \bibinfo
  {pages} {200501} (\bibinfo {year} {2018})}\BibitemShut {NoStop}%
\bibitem [{\citenamefont {Kurpiers}\ \emph {et~al.}(2018)\citenamefont
  {Kurpiers}, \citenamefont {Magnard}, \citenamefont {Walter}, \citenamefont
  {Royer}, \citenamefont {Pechal}, \citenamefont {Heinsoo}, \citenamefont
  {Salath{\'e}}, \citenamefont {Akin}, \citenamefont {Storz}, \citenamefont
  {Besse}, \citenamefont {Gasparinetti}, \citenamefont {Blais},\ and\
  \citenamefont {Wallraff}}]{kurpiersDeterministicQuantumState2018}%
  \BibitemOpen
  \bibfield  {author} {\bibinfo {author} {\bibfnamefont {P.}~\bibnamefont
  {Kurpiers}}, \bibinfo {author} {\bibfnamefont {P.}~\bibnamefont {Magnard}},
  \bibinfo {author} {\bibfnamefont {T.}~\bibnamefont {Walter}}, \bibinfo
  {author} {\bibfnamefont {B.}~\bibnamefont {Royer}}, \bibinfo {author}
  {\bibfnamefont {M.}~\bibnamefont {Pechal}}, \bibinfo {author} {\bibfnamefont
  {J.}~\bibnamefont {Heinsoo}}, \bibinfo {author} {\bibfnamefont
  {Y.}~\bibnamefont {Salath{\'e}}}, \bibinfo {author} {\bibfnamefont
  {A.}~\bibnamefont {Akin}}, \bibinfo {author} {\bibfnamefont {S.}~\bibnamefont
  {Storz}}, \bibinfo {author} {\bibfnamefont {J.-C.}\ \bibnamefont {Besse}},
  \bibinfo {author} {\bibfnamefont {S.}~\bibnamefont {Gasparinetti}}, \bibinfo
  {author} {\bibfnamefont {A.}~\bibnamefont {Blais}}, \ and\ \bibinfo {author}
  {\bibfnamefont {A.}~\bibnamefont {Wallraff}},\ }\bibfield  {title} {\enquote
  {\bibinfo {title} {Deterministic quantum state transfer and remote
  entanglement using microwave photons},}\ }\href {\doibase
  10.1038/s41586-018-0195-y} {\bibfield  {journal} {\bibinfo  {journal}
  {Nature}\ }\textbf {\bibinfo {volume} {558}},\ \bibinfo {pages} {264}
  (\bibinfo {year} {2018})}\BibitemShut {NoStop}%
\bibitem [{\citenamefont {Axline}\ \emph {et~al.}(2018)\citenamefont {Axline},
  \citenamefont {Burkhart}, \citenamefont {Pfaff}, \citenamefont {Zhang},
  \citenamefont {Chou}, \citenamefont {{Campagne-Ibarcq}}, \citenamefont
  {Reinhold}, \citenamefont {Frunzio}, \citenamefont {Girvin}, \citenamefont
  {Jiang}, \citenamefont {Devoret},\ and\ \citenamefont
  {Schoelkopf}}]{axlineOndemandQuantumState2018}%
  \BibitemOpen
  \bibfield  {author} {\bibinfo {author} {\bibfnamefont {Christopher~J.}\
  \bibnamefont {Axline}}, \bibinfo {author} {\bibfnamefont {Luke~D.}\
  \bibnamefont {Burkhart}}, \bibinfo {author} {\bibfnamefont {Wolfgang}\
  \bibnamefont {Pfaff}}, \bibinfo {author} {\bibfnamefont {Mengzhen}\
  \bibnamefont {Zhang}}, \bibinfo {author} {\bibfnamefont {Kevin}\ \bibnamefont
  {Chou}}, \bibinfo {author} {\bibfnamefont {Philippe}\ \bibnamefont
  {{Campagne-Ibarcq}}}, \bibinfo {author} {\bibfnamefont {Philip}\ \bibnamefont
  {Reinhold}}, \bibinfo {author} {\bibfnamefont {Luigi}\ \bibnamefont
  {Frunzio}}, \bibinfo {author} {\bibfnamefont {S.~M.}\ \bibnamefont {Girvin}},
  \bibinfo {author} {\bibfnamefont {Liang}\ \bibnamefont {Jiang}}, \bibinfo
  {author} {\bibfnamefont {M.~H.}\ \bibnamefont {Devoret}}, \ and\ \bibinfo
  {author} {\bibfnamefont {R.~J.}\ \bibnamefont {Schoelkopf}},\ }\bibfield
  {title} {\enquote {\bibinfo {title} {On-demand quantum state transfer and
  entanglement between remote microwave cavity memories},}\ }\href {\doibase
  10.1038/s41567-018-0115-y} {\bibfield  {journal} {\bibinfo  {journal} {Nature
  Physics}\ }\textbf {\bibinfo {volume} {14}},\ \bibinfo {pages} {705}
  (\bibinfo {year} {2018})}\BibitemShut {NoStop}%
\bibitem [{\citenamefont {Franson}(1989)}]{fransonBellInequalityPosition1989}%
  \BibitemOpen
  \bibfield  {author} {\bibinfo {author} {\bibfnamefont {J.~D.}\ \bibnamefont
  {Franson}},\ }\bibfield  {title} {\enquote {\bibinfo {title} {Bell inequality
  for position and time},}\ }\href {\doibase 10.1103/PhysRevLett.62.2205}
  {\bibfield  {journal} {\bibinfo  {journal} {Physical Review Letters}\
  }\textbf {\bibinfo {volume} {62}},\ \bibinfo {pages} {2205--2208} (\bibinfo
  {year} {1989})}\BibitemShut {NoStop}%
\bibitem [{\citenamefont {Rarity}\ \emph {et~al.}(1990)\citenamefont {Rarity},
  \citenamefont {Tapster}, \citenamefont {Jakeman}, \citenamefont {Larchuk},
  \citenamefont {Campos}, \citenamefont {Teich},\ and\ \citenamefont
  {Saleh}}]{rarityTwophotonInterferenceMachZehnder1990}%
  \BibitemOpen
  \bibfield  {author} {\bibinfo {author} {\bibfnamefont {J.~G.}\ \bibnamefont
  {Rarity}}, \bibinfo {author} {\bibfnamefont {P.~R.}\ \bibnamefont {Tapster}},
  \bibinfo {author} {\bibfnamefont {E.}~\bibnamefont {Jakeman}}, \bibinfo
  {author} {\bibfnamefont {T.}~\bibnamefont {Larchuk}}, \bibinfo {author}
  {\bibfnamefont {R.~A.}\ \bibnamefont {Campos}}, \bibinfo {author}
  {\bibfnamefont {M.~C.}\ \bibnamefont {Teich}}, \ and\ \bibinfo {author}
  {\bibfnamefont {B.~E.~A.}\ \bibnamefont {Saleh}},\ }\bibfield  {title}
  {\enquote {\bibinfo {title} {Two-photon interference in a
  {{Mach}}-{{Zehnder}} interferometer},}\ }\href {\doibase
  10.1103/PhysRevLett.65.1348} {\bibfield  {journal} {\bibinfo  {journal}
  {Physical Review Letters}\ }\textbf {\bibinfo {volume} {65}},\ \bibinfo
  {pages} {1348--1351} (\bibinfo {year} {1990})}\BibitemShut {NoStop}%
\bibitem [{\citenamefont {Bernien}\ \emph {et~al.}(2013)\citenamefont
  {Bernien}, \citenamefont {Hensen}, \citenamefont {Pfaff}, \citenamefont
  {Koolstra}, \citenamefont {Blok}, \citenamefont {Robledo}, \citenamefont
  {Taminiau}, \citenamefont {Markham}, \citenamefont {Twitchen},\ and\
  \citenamefont {Hanson}}]{bernienHeraldedEntanglementSolidstate2013}%
  \BibitemOpen
  \bibfield  {author} {\bibinfo {author} {\bibfnamefont {H.}~\bibnamefont
  {Bernien}}, \bibinfo {author} {\bibfnamefont {B.}~\bibnamefont {Hensen}},
  \bibinfo {author} {\bibfnamefont {W.}~\bibnamefont {Pfaff}}, \bibinfo
  {author} {\bibfnamefont {G.}~\bibnamefont {Koolstra}}, \bibinfo {author}
  {\bibfnamefont {M.~S.}\ \bibnamefont {Blok}}, \bibinfo {author}
  {\bibfnamefont {L.}~\bibnamefont {Robledo}}, \bibinfo {author} {\bibfnamefont
  {T.~H.}\ \bibnamefont {Taminiau}}, \bibinfo {author} {\bibfnamefont
  {M.}~\bibnamefont {Markham}}, \bibinfo {author} {\bibfnamefont {D.~J.}\
  \bibnamefont {Twitchen}}, \ and\ \bibinfo {author} {\bibfnamefont
  {R.}~\bibnamefont {Hanson}},\ }\bibfield  {title} {\enquote {\bibinfo {title}
  {Heralded entanglement between solid-state qubits separated by three
  metres},}\ }\href@noop {} {\bibfield  {journal} {\bibinfo  {journal}
  {Nature}\ }\textbf {\bibinfo {volume} {497}},\ \bibinfo {pages} {86}
  (\bibinfo {year} {2013})}\BibitemShut {NoStop}%
\bibitem [{\citenamefont {Kurpiers}\ \emph {et~al.}(2019)\citenamefont
  {Kurpiers}, \citenamefont {Pechal}, \citenamefont {Royer}, \citenamefont
  {Magnard}, \citenamefont {Walter}, \citenamefont {Heinsoo}, \citenamefont
  {Salath{\'e}}, \citenamefont {Akin}, \citenamefont {Storz}, \citenamefont
  {Besse}, \citenamefont {Gasparinetti}, \citenamefont {Blais},\ and\
  \citenamefont {Wallraff}}]{kurpiersQuantumCommunicationTimebin2019}%
  \BibitemOpen
  \bibfield  {author} {\bibinfo {author} {\bibfnamefont {P.}~\bibnamefont
  {Kurpiers}}, \bibinfo {author} {\bibfnamefont {M.}~\bibnamefont {Pechal}},
  \bibinfo {author} {\bibfnamefont {B.}~\bibnamefont {Royer}}, \bibinfo
  {author} {\bibfnamefont {P.}~\bibnamefont {Magnard}}, \bibinfo {author}
  {\bibfnamefont {T.}~\bibnamefont {Walter}}, \bibinfo {author} {\bibfnamefont
  {J.}~\bibnamefont {Heinsoo}}, \bibinfo {author} {\bibfnamefont
  {Y.}~\bibnamefont {Salath{\'e}}}, \bibinfo {author} {\bibfnamefont
  {A.}~\bibnamefont {Akin}}, \bibinfo {author} {\bibfnamefont {S.}~\bibnamefont
  {Storz}}, \bibinfo {author} {\bibfnamefont {J.-C.}\ \bibnamefont {Besse}},
  \bibinfo {author} {\bibfnamefont {S.}~\bibnamefont {Gasparinetti}}, \bibinfo
  {author} {\bibfnamefont {A.}~\bibnamefont {Blais}}, \ and\ \bibinfo {author}
  {\bibfnamefont {A.}~\bibnamefont {Wallraff}},\ }\bibfield  {title} {\enquote
  {\bibinfo {title} {Quantum {{Communication}} with {{Time}}-{{Bin Encoded
  Microwave Photons}}},}\ }\href {\doibase 10.1103/PhysRevApplied.12.044067}
  {\bibfield  {journal} {\bibinfo  {journal} {Physical Review Applied}\
  }\textbf {\bibinfo {volume} {12}},\ \bibinfo {pages} {044067} (\bibinfo
  {year} {2019})}\BibitemShut {NoStop}%
\end{thebibliography}

\begin{thebibliography}{10}%
\makeatletter
\providecommand \@ifxundefined [1]{%
 \@ifx{#1\undefined}
}%
\providecommand \@ifnum [1]{%
 \ifnum #1\expandafter \@firstoftwo
 \else \expandafter \@secondoftwo
 \fi
}%
\providecommand \@ifx [1]{%
 \ifx #1\expandafter \@firstoftwo
 \else \expandafter \@secondoftwo
 \fi
}%
\providecommand \natexlab [1]{#1}%
\providecommand \enquote  [1]{``#1''}%
\providecommand \bibnamefont  [1]{#1}%
\providecommand \bibfnamefont [1]{#1}%
\providecommand \citenamefont [1]{#1}%
\providecommand \href@noop [0]{\@secondoftwo}%
\providecommand \href [0]{\begingroup \@sanitize@url \@href}%
\providecommand \@href[1]{\@@startlink{#1}\@@href}%
\providecommand \@@href[1]{\endgroup#1\@@endlink}%
\providecommand \@sanitize@url [0]{\catcode `\\12\catcode `\$12\catcode
  `\&12\catcode `\#12\catcode `\^12\catcode `\_12\catcode `\%12\relax}%
\providecommand \@@startlink[1]{}%
\providecommand \@@endlink[0]{}%
\providecommand \url  [0]{\begingroup\@sanitize@url \@url }%
\providecommand \@url [1]{\endgroup\@href {#1}{\urlprefix }}%
\providecommand \urlprefix  [0]{URL }%
\providecommand \Eprint [0]{\href }%
\providecommand \doibase [0]{http://dx.doi.org/}%
\providecommand \selectlanguage [0]{\@gobble}%
\providecommand \bibinfo  [0]{\@secondoftwo}%
\providecommand \bibfield  [0]{\@secondoftwo}%
\providecommand \translation [1]{[#1]}%
\providecommand \BibitemOpen [0]{}%
\providecommand \bibitemStop [0]{}%
\providecommand \bibitemNoStop [0]{.\EOS\space}%
\providecommand \EOS [0]{\spacefactor3000\relax}%
\providecommand \BibitemShut  [1]{\csname bibitem#1\endcsname}%
\let\auto@bib@innerbib\@empty
%</preamble>
\bibitem [{\citenamefont {Bienfait}\ \emph {et~al.}(2019)\citenamefont
  {Bienfait}, \citenamefont {Satzinger}, \citenamefont {Zhong}, \citenamefont
  {Chang}, \citenamefont {Chou}, \citenamefont {Conner}, \citenamefont {Dumur},
  \citenamefont {Grebel}, \citenamefont {Peairs}, \citenamefont {Povey},\ and\
  \citenamefont {Cleland}}]{bienfaitPhononmediatedQuantumState2019}%
  \BibitemOpen
  \bibfield  {author} {\bibinfo {author} {\bibfnamefont {A.}~\bibnamefont
  {Bienfait}}, \bibinfo {author} {\bibfnamefont {K.~J.}\ \bibnamefont
  {Satzinger}}, \bibinfo {author} {\bibfnamefont {Y.~P.}\ \bibnamefont
  {Zhong}}, \bibinfo {author} {\bibfnamefont {H.-S.}\ \bibnamefont {Chang}},
  \bibinfo {author} {\bibfnamefont {M.-H.}\ \bibnamefont {Chou}}, \bibinfo
  {author} {\bibfnamefont {C.~R.}\ \bibnamefont {Conner}}, \bibinfo {author}
  {\bibfnamefont {{\'E}.}~\bibnamefont {Dumur}}, \bibinfo {author}
  {\bibfnamefont {J.}~\bibnamefont {Grebel}}, \bibinfo {author} {\bibfnamefont
  {G.~A.}\ \bibnamefont {Peairs}}, \bibinfo {author} {\bibfnamefont {R.~G.}\
  \bibnamefont {Povey}}, \ and\ \bibinfo {author} {\bibfnamefont {A.~N.}\
  \bibnamefont {Cleland}},\ }\href {\doibase 10.1126/science.aaw8415}
  {\bibfield  {journal} {\bibinfo  {journal} {Science}\ }\textbf {\bibinfo
  {volume} {364}},\ \bibinfo {pages} {368} (\bibinfo {year}
  {2019})}\BibitemShut {NoStop}%
\bibitem [{\citenamefont {Zhong}\ \emph {et~al.}(2019)\citenamefont {Zhong},
  \citenamefont {Chang}, \citenamefont {Satzinger}, \citenamefont {Chou},
  \citenamefont {Bienfait}, \citenamefont {Conner}, \citenamefont {Dumur},
  \citenamefont {Grebel}, \citenamefont {Peairs}, \citenamefont {Povey},
  \citenamefont {Schuster},\ and\ \citenamefont
  {Cleland}}]{zhongViolatingBellInequality2018}%
  \BibitemOpen
  \bibfield  {author} {\bibinfo {author} {\bibfnamefont {Y.~P.}\ \bibnamefont
  {Zhong}}, \bibinfo {author} {\bibfnamefont {H.-S.}\ \bibnamefont {Chang}},
  \bibinfo {author} {\bibfnamefont {K.~J.}\ \bibnamefont {Satzinger}}, \bibinfo
  {author} {\bibfnamefont {M.-H.}\ \bibnamefont {Chou}}, \bibinfo {author}
  {\bibfnamefont {A.}~\bibnamefont {Bienfait}}, \bibinfo {author}
  {\bibfnamefont {C.~R.}\ \bibnamefont {Conner}}, \bibinfo {author}
  {\bibfnamefont {{\'E}.}~\bibnamefont {Dumur}}, \bibinfo {author}
  {\bibfnamefont {J.}~\bibnamefont {Grebel}}, \bibinfo {author} {\bibfnamefont
  {G.~A.}\ \bibnamefont {Peairs}}, \bibinfo {author} {\bibfnamefont {R.~G.}\
  \bibnamefont {Povey}}, \bibinfo {author} {\bibfnamefont {D.~I.}\ \bibnamefont
  {Schuster}}, \ and\ \bibinfo {author} {\bibfnamefont {A.~N.}\ \bibnamefont
  {Cleland}},\ }\href {\doibase 10.1038/s41567-019-0507-7} {\bibfield
  {journal} {\bibinfo  {journal} {Nature Physics}\ }\textbf {\bibinfo {volume}
  {15}},\ \bibinfo {pages} {741} (\bibinfo {year} {2019})}\BibitemShut
  {NoStop}%
\bibitem [{\citenamefont {Satzinger}\ \emph {et~al.}(2018)\citenamefont
  {Satzinger}, \citenamefont {Zhong}, \citenamefont {Chang}, \citenamefont
  {Peairs}, \citenamefont {Bienfait}, \citenamefont {Chou}, \citenamefont
  {Cleland}, \citenamefont {Conner}, \citenamefont {Dumur}, \citenamefont
  {Grebel}, \citenamefont {Gutierrez}, \citenamefont {November}, \citenamefont
  {Povey}, \citenamefont {Whiteley}, \citenamefont {Awschalom}, \citenamefont
  {Schuster},\ and\ \citenamefont
  {Cleland}}]{satzingerQuantumControlSurface2018}%
  \BibitemOpen
  \bibfield  {author} {\bibinfo {author} {\bibfnamefont {K.~J.}\ \bibnamefont
  {Satzinger}}, \bibinfo {author} {\bibfnamefont {Y.~P.}\ \bibnamefont
  {Zhong}}, \bibinfo {author} {\bibfnamefont {H.-S.}\ \bibnamefont {Chang}},
  \bibinfo {author} {\bibfnamefont {G.~A.}\ \bibnamefont {Peairs}}, \bibinfo
  {author} {\bibfnamefont {A.}~\bibnamefont {Bienfait}}, \bibinfo {author}
  {\bibfnamefont {M.-H.}\ \bibnamefont {Chou}}, \bibinfo {author}
  {\bibfnamefont {A.~Y.}\ \bibnamefont {Cleland}}, \bibinfo {author}
  {\bibfnamefont {C.~R.}\ \bibnamefont {Conner}}, \bibinfo {author}
  {\bibfnamefont {{\'E}.}~\bibnamefont {Dumur}}, \bibinfo {author}
  {\bibfnamefont {J.}~\bibnamefont {Grebel}}, \bibinfo {author} {\bibfnamefont
  {I.}~\bibnamefont {Gutierrez}}, \bibinfo {author} {\bibfnamefont {B.~H.}\
  \bibnamefont {November}}, \bibinfo {author} {\bibfnamefont {R.~G.}\
  \bibnamefont {Povey}}, \bibinfo {author} {\bibfnamefont {S.~J.}\ \bibnamefont
  {Whiteley}}, \bibinfo {author} {\bibfnamefont {D.~D.}\ \bibnamefont
  {Awschalom}}, \bibinfo {author} {\bibfnamefont {D.~I.}\ \bibnamefont
  {Schuster}}, \ and\ \bibinfo {author} {\bibfnamefont {A.~N.}\ \bibnamefont
  {Cleland}},\ }\href {\doibase 10.1038/s41586-018-0719-5} {\bibfield
  {journal} {\bibinfo  {journal} {Nature}\ }\textbf {\bibinfo {volume} {563}},\
  \bibinfo {pages} {661} (\bibinfo {year} {2018})}\BibitemShut {NoStop}%
\bibitem [{\citenamefont {Macklin}\ \emph {et~al.}(2015)\citenamefont
  {Macklin}, \citenamefont {O'Brien}, \citenamefont {Hover}, \citenamefont
  {Schwartz}, \citenamefont {Bolkhovsky}, \citenamefont {Zhang}, \citenamefont
  {Oliver},\ and\ \citenamefont
  {Siddiqi}}]{macklinQuantumlimitedJosephsonTravelingwave2015}%
  \BibitemOpen
  \bibfield  {author} {\bibinfo {author} {\bibfnamefont {C.}~\bibnamefont
  {Macklin}}, \bibinfo {author} {\bibfnamefont {K.}~\bibnamefont {O'Brien}},
  \bibinfo {author} {\bibfnamefont {D.}~\bibnamefont {Hover}}, \bibinfo
  {author} {\bibfnamefont {M.~E.}\ \bibnamefont {Schwartz}}, \bibinfo {author}
  {\bibfnamefont {V.}~\bibnamefont {Bolkhovsky}}, \bibinfo {author}
  {\bibfnamefont {X.}~\bibnamefont {Zhang}}, \bibinfo {author} {\bibfnamefont
  {W.~D.}\ \bibnamefont {Oliver}}, \ and\ \bibinfo {author} {\bibfnamefont
  {I.}~\bibnamefont {Siddiqi}},\ }\href {\doibase 10.1126/science.aaa8525}
  {\bibfield  {journal} {\bibinfo  {journal} {Science}\ }\textbf {\bibinfo
  {volume} {350}},\ \bibinfo {pages} {307} (\bibinfo {year}
  {2015})}\BibitemShut {NoStop}%
\bibitem [{\citenamefont {Koch}\ \emph {et~al.}(2007)\citenamefont {Koch},
  \citenamefont {Yu}, \citenamefont {Gambetta}, \citenamefont {Houck},
  \citenamefont {Schuster}, \citenamefont {Majer}, \citenamefont {Blais},
  \citenamefont {Devoret}, \citenamefont {Girvin},\ and\ \citenamefont
  {Schoelkopf}}]{kochChargeinsensitiveQubitDesign2007}%
  \BibitemOpen
  \bibfield  {author} {\bibinfo {author} {\bibfnamefont {J.}~\bibnamefont
  {Koch}}, \bibinfo {author} {\bibfnamefont {T.~M.}\ \bibnamefont {Yu}},
  \bibinfo {author} {\bibfnamefont {J.}~\bibnamefont {Gambetta}}, \bibinfo
  {author} {\bibfnamefont {A.~A.}\ \bibnamefont {Houck}}, \bibinfo {author}
  {\bibfnamefont {D.~I.}\ \bibnamefont {Schuster}}, \bibinfo {author}
  {\bibfnamefont {J.}~\bibnamefont {Majer}}, \bibinfo {author} {\bibfnamefont
  {A.}~\bibnamefont {Blais}}, \bibinfo {author} {\bibfnamefont {M.~H.}\
  \bibnamefont {Devoret}}, \bibinfo {author} {\bibfnamefont {S.~M.}\
  \bibnamefont {Girvin}}, \ and\ \bibinfo {author} {\bibfnamefont {R.~J.}\
  \bibnamefont {Schoelkopf}},\ }\href {\doibase 10.1103/PhysRevA.76.042319}
  {\bibfield  {journal} {\bibinfo  {journal} {Phys. Rev. A}\ }\textbf {\bibinfo
  {volume} {76}},\ \bibinfo {pages} {042319} (\bibinfo {year}
  {2007})}\BibitemShut {NoStop}%
\bibitem [{\citenamefont {Frisk~Kockum}\ \emph {et~al.}(2014)\citenamefont
  {Frisk~Kockum}, \citenamefont {Delsing},\ and\ \citenamefont
  {Johansson}}]{friskkockumDesigningFrequencydependentRelaxation2014}%
  \BibitemOpen
  \bibfield  {author} {\bibinfo {author} {\bibfnamefont {A.}~\bibnamefont
  {Frisk~Kockum}}, \bibinfo {author} {\bibfnamefont {P.}~\bibnamefont
  {Delsing}}, \ and\ \bibinfo {author} {\bibfnamefont {G.}~\bibnamefont
  {Johansson}},\ }\href {\doibase 10.1103/PhysRevA.90.013837} {\bibfield
  {journal} {\bibinfo  {journal} {Physical Review A}\ }\textbf {\bibinfo
  {volume} {90}},\ \bibinfo {pages} {013837} (\bibinfo {year}
  {2014})}\BibitemShut {NoStop}%
\bibitem [{\citenamefont {Morgan}(2007)}]{morgandavidSurfaceAcousticWave2007}%
  \BibitemOpen
  \bibfield  {author} {\bibinfo {author} {\bibfnamefont {D.}~\bibnamefont
  {Morgan}},\ }\href {\doibase 10.1016/B978-0-12-372537-0.X5000-6} {\emph
  {\bibinfo {title} {Surface {{Acoustic Wave Filters}}}}}\ (\bibinfo
  {publisher} {{Elsevier}},\ \bibinfo {year} {2007})\BibitemShut {NoStop}%
\bibitem [{\citenamefont {Chen}\ \emph {et~al.}(2014)\citenamefont {Chen},
  \citenamefont {Neill}, \citenamefont {Roushan}, \citenamefont {Leung},
  \citenamefont {Fang}, \citenamefont {Barends}, \citenamefont {Kelly},
  \citenamefont {Campbell}, \citenamefont {Chen}, \citenamefont {Chiaro},
  \citenamefont {Dunsworth}, \citenamefont {Jeffrey}, \citenamefont {Megrant},
  \citenamefont {Mutus}, \citenamefont {O'Malley}, \citenamefont {Quintana},
  \citenamefont {Sank}, \citenamefont {Vainsencher}, \citenamefont {Wenner},
  \citenamefont {White}, \citenamefont {Geller}, \citenamefont {Cleland},\ and\
  \citenamefont {Martinis}}]{chenQubitArchitectureHigh2014}%
  \BibitemOpen
  \bibfield  {author} {\bibinfo {author} {\bibfnamefont {Y.}~\bibnamefont
  {Chen}}, \bibinfo {author} {\bibfnamefont {C.}~\bibnamefont {Neill}},
  \bibinfo {author} {\bibfnamefont {P.}~\bibnamefont {Roushan}}, \bibinfo
  {author} {\bibfnamefont {N.}~\bibnamefont {Leung}}, \bibinfo {author}
  {\bibfnamefont {M.}~\bibnamefont {Fang}}, \bibinfo {author} {\bibfnamefont
  {R.}~\bibnamefont {Barends}}, \bibinfo {author} {\bibfnamefont
  {J.}~\bibnamefont {Kelly}}, \bibinfo {author} {\bibfnamefont
  {B.}~\bibnamefont {Campbell}}, \bibinfo {author} {\bibfnamefont
  {Z.}~\bibnamefont {Chen}}, \bibinfo {author} {\bibfnamefont {B.}~\bibnamefont
  {Chiaro}}, \bibinfo {author} {\bibfnamefont {A.}~\bibnamefont {Dunsworth}},
  \bibinfo {author} {\bibfnamefont {E.}~\bibnamefont {Jeffrey}}, \bibinfo
  {author} {\bibfnamefont {A.}~\bibnamefont {Megrant}}, \bibinfo {author}
  {\bibfnamefont {J.~Y.}\ \bibnamefont {Mutus}}, \bibinfo {author}
  {\bibfnamefont {P.~J.~J.}\ \bibnamefont {O'Malley}}, \bibinfo {author}
  {\bibfnamefont {C.~M.}\ \bibnamefont {Quintana}}, \bibinfo {author}
  {\bibfnamefont {D.}~\bibnamefont {Sank}}, \bibinfo {author} {\bibfnamefont
  {A.}~\bibnamefont {Vainsencher}}, \bibinfo {author} {\bibfnamefont
  {J.}~\bibnamefont {Wenner}}, \bibinfo {author} {\bibfnamefont {T.~C.}\
  \bibnamefont {White}}, \bibinfo {author} {\bibfnamefont {M.~R.}\ \bibnamefont
  {Geller}}, \bibinfo {author} {\bibfnamefont {A.~N.}\ \bibnamefont {Cleland}},
  \ and\ \bibinfo {author} {\bibfnamefont {J.~M.}\ \bibnamefont {Martinis}},\
  }\href {\doibase 10.1103/PhysRevLett.113.220502} {\bibfield  {journal}
  {\bibinfo  {journal} {Physical Review Letters}\ }\textbf {\bibinfo {volume}
  {113}},\ \bibinfo {pages} {220502} (\bibinfo {year} {2014})}\BibitemShut
  {NoStop}%
\bibitem [{\citenamefont {Kiilerich}\ and\ \citenamefont
  {M{\o}lmer}(2019)}]{kiilerichInputOutputTheoryQuantum2019}%
  \BibitemOpen
  \bibfield  {author} {\bibinfo {author} {\bibfnamefont {A.~H.}\ \bibnamefont
  {Kiilerich}}\ and\ \bibinfo {author} {\bibfnamefont {K.}~\bibnamefont
  {M{\o}lmer}},\ }\href {\doibase 10.1103/PhysRevLett.123.123604} {\bibfield
  {journal} {\bibinfo  {journal} {Physical Review Letters}\ }\textbf {\bibinfo
  {volume} {123}},\ \bibinfo {pages} {123604} (\bibinfo {year}
  {2019})}\BibitemShut {NoStop}%
\bibitem [{\citenamefont {Johansson}\ \emph {et~al.}(2012)\citenamefont
  {Johansson}, \citenamefont {Nation},\ and\ \citenamefont
  {Nori}}]{johanssonQuTiPOpensourcePython2012}%
  \BibitemOpen
  \bibfield  {author} {\bibinfo {author} {\bibfnamefont {J.~R.}\ \bibnamefont
  {Johansson}}, \bibinfo {author} {\bibfnamefont {P.~D.}\ \bibnamefont
  {Nation}}, \ and\ \bibinfo {author} {\bibfnamefont {F.}~\bibnamefont
  {Nori}},\ }\href {\doibase 10.1016/j.cpc.2012.02.021} {\bibfield  {journal}
  {\bibinfo  {journal} {Computer Physics Communications}\ }\textbf {\bibinfo
  {volume} {183}},\ \bibinfo {pages} {1760} (\bibinfo {year}
  {2012})}\BibitemShut {NoStop}%
\end{thebibliography}
\end{document}